\documentclass{aa}  

\newcommand{\kms}{km\ s$^{-1}$}

\newcommand{\CHX}{CHX~22}
\newcommand{\HPCHA}{HP~Cha}
\newcommand{\SCRA}{S~CrA}

\usepackage{hyperref}
\usepackage{color}
\usepackage{graphicx}
\usepackage{txfonts}
\begin{document} 

   \title{Disk Evolution Study Through Imaging of Nearby Young Stars (DESTINYS): Diverse outcomes of binary-disk interactions}

   \subtitle{}

    \author{Yapeng Zhang \inst{1}
     \and Christian Ginski \inst{1,2}
     \and Jane Huang \inst{3,19}
     \and Alice Zurlo\inst{4,5}
     \and Herv\'e Beust \inst{6}
     \and Jaehan Bae \inst{7}
     \and Myriam Benisty\inst{8,6}
     \and Antonio Garufi\inst{9}
     \and Michiel R. Hogerheijde\inst{1,2}
     \and Rob G. van Holstein\inst{10}
     \and Matthew Kenworthy\inst{1}
     \and Maud Langlois\inst{11}
     \and Carlo F. Manara\inst{12}
     \and Paola Pinilla\inst{13}
     \and Christian Rab\inst{14,15}
     \and \'Alvaro Ribas\inst{16}
     \and Giovanni P. Rosotti\inst{17,1}
     \and Jonathan Williams \inst{18}
    }
        
   \institute{Leiden Observatory, Leiden University, Postbus 9513, 2300 RA, Leiden, The Netherlands \\
   \email{yzhang@strw.leidenuniv.nl}
   \and Anton Pannekoek Institute for Astronomy, University of Amsterdam, Science Park 904, 1098 XH Amsterdam, The Netherlands
   \and Department of Astronomy, University of Michigan, 323 West Hall, 1085 S. University Avenue, Ann Arbor, MI 48109, USA
   \and Escuela de Ingenier\'ia Industrial, Facultad de Ingenier\'ia y Ciencias, Universidad Diego Portales, Av. Ejercito 441, Santiago, Chile
   \and Millennium Nucleus on Young Exoplanets and their Moons (YEMS)
   \and Univ. Grenoble Alpes, CNRS, IPAG, 38000 Grenoble, France
   \and Department of Astronomy, University of Florida, Gainesville, FL 32611, USA
   \and Laboratoire Lagrange, Université Côte d’Azur, CNRS, Observatoire de la Côte d’Azur, 06304 Nice, France
   \and INAF, Osservatorio Astrofisico di Arcetri, Largo Enrico Fermi 5, I-50125 Firenze, Italy
   \and European Southern Observatory, Alonso de Córdova 3107, Vitacura, Casilla 19001, Santiago de Chile, Chile
   \and Centre de Recherche Astrophysique de Lyon 9 Ave. Charles André 69561 Saint-Genis Laval, France
   \and European Southern Observatory, Karl-Schwarzschild-Strasse 2, 85748 Garching bei München, Germany
   \and Mullard Space Science Laboratory, University College London, Holmbury St Mary, Dorking, UK
   \and University Observatory, Faculty of Physics, Ludwig-Maximilians-Universität München, Scheinerstr. 1, 81679 Munich, Germany
   \and Max-Planck-Institut f\"ur extraterrestrische Physik, Giessenbachstrasse 1, 85748 Garching, Germany
   \and Institute of Astronomy, University of Cambridge, Madingley Road, Cambridge CB3 0HA, UK
   \and Dipartimento di Fisica, Università degli Studi di Milano via Giovanni Celoria 16, 20133 Milano, Italy
   \and Institute for Astronomy, University of Hawai'i at Mānoa, Honolulu, HI 96822, USA
   \and NASA Hubble Fellowship Program Sagan Fellow
   }
    \authorrunning{Y. Zhang et al.}
    \titlerunning{DESTINYS: Binary-Disk}

  \date{Received November 30, 2022; accepted February 24, 2023}

% 5 {} token are mandatory: context, aims, methods, results, conclusion.
\abstract{Circumstellar disks play an essential role in the outcomes of planet formation. Disks do not evolve in isolation, as about half of solar-type stars were born in binary or multiple systems. The presence of stellar companions modifies the morphology and evolution of disks, potentially resulting in a different planet population. Resolving disks in binary systems provides the opportunity to examine the influence of stellar companions on the outcomes of planet formation.
}
{We aim to investigate and compare disks in stellar multiple systems with near-infrared scattered-light imaging as part of the Disk Evolution Study Through Imaging of Nearby Young Stars (DESTINYS) large program. In particular, we present observations of circumstellar disks in three systems, namely, \CHX, \SCRA, and \HPCHA.
}
{We used polarimetric differential imaging with SPHERE/IRDIS at the VLT to search for scattered light from the circumstellar disks in these multiple systems. We performed astrometric and orbit analyses for the stellar companions using archival HST, VLT/NACO, and SPHERE data to better understand the interplay between disks and companions. }
{Combined with the age and orbital constraints, the observed disk structures in scattered light provide insights into the evolutionary history and the impact of the stellar companions. The small grains in \CHX\ form a tail-like structure surrounding the close binary, which likely results from a close encounter and capture of a cloudlet. \SCRA\ shows intricate structures (tentative ringed and spiral features) in the circumprimary disk as a possible consequence of perturbations by companions. The circumsecondary disk is truncated and connected to the primary disk via a streamer, suggesting tidal interactions. In \HPCHA, the primary disk is less disturbed and features a tenuous streamer, through which the material flows towards the companions. }
{The comparison of the three systems spans a wide range of binary separation ($50 - 500$ au) and illustrates the decreasing influence on disk structures with the distance of companions. This agrees with the statistical analysis of exoplanet population in binaries, that planet formation is likely obstructed around close binary systems, while it is not suppressed in wide binaries.} 


   \maketitle
%
%-------------------------------------------------------------------

\section{Introduction}

Circumstellar disks as the birthplaces of planets play an important role in the outcomes of planet formation. Dust in the disks undergoing growth into pebbles and planetesimals constitute the building blocks of planetary embryos. Therefore the properties of dust and their evolution significantly influence the planet formation processes \citep{Drazkowska2022}. In turn, gas giants formed in disks affect the distribution of solids and disk structures. 
Recent high angular resolution observations of circumstellar disks across a wide wavelength range reveal near ubiquitous substructures (rings, spirals, and cavities) which may be carved by newborn planets \citep{Andrews2018, Bae2018, Lodato2019, Benisty2022}. Therefore understanding the interplay between disks and massive objects is essential for unraveling planet formation.

Studies of circumstellar disks have been mostly focused on isolated stars. However, it is well established that about half of solar-type stars were born in  multiple systems \citep{Duquennoy1991, Raghavan2010}, hence the potential effects of binarity on disk evolution and planet formation are not negligible. In particular, the gravitational perturbation and tidal truncation due to the close central binaries or outer companions will immensely impact the morphology of the disks, such as opening large inner cavities, truncating disk sizes, inducing spiral features, or warping and misaligning inner and outer disk regions \citep{Papaloizou1977, Artymowicz1994, Marino2015, Benisty2017, Rosotti2018, Ragusa2020}. The presence of companions may also dictate the evolution of gas and dust in disks \citep{Rosotti2018, Ribas2018, Zagaria2021a, Zagaria2021b}.
Dedicated surveys of star-forming regions suggest that protoplanetary disks in multiple systems are more compact and less massive than disks around isolated stars \citep{Cox2017, Manara2019, Akeson2019, Zurlo2020, Zurlo2021, Rota2022}.

Despite the dominance of destructive processes around multiple stars, several disks in such systems have been previously imaged in scattered light, such as GG Tau \citep{Keppler2020}, GW Ori \citep{Kraus2020}, and UX Tau \citep{Zapata2020, Menard2020}, for which detailed hydrodynamical simulations have been dedicated in order to provide insights into the influence of stellar companions on disk structures.
Observations and simulations of disks in binaries provide opportunities to examine the plausibility of planet formation in various conditions, which can be further linked to the demographics of exoplanets in  multiple stellar systems. 

Among the five-thousand exoplanets discovered to date, more than two hundred were discovered in multiple systems \citep{Schwarz2016, Martin2018}, suggesting that planet formation in such systems indeed occurs. Most of these planets are found surrounding individual components of the binaries (called S-type architecture) as opposed to circumbinary planets (called P-type).
Multiplicity surveys of planet-hosting stars have been carried out using direct imaging \citep[e.g.,][]{Eggenberger2007, Mugrauer2015, Mugrauer2020, Ngo2015, Ginski2012, Ginski2016, Ginski2021a, Bohn2020}.
The statistical analyses of the impact of binaries on planet frequencies remain inconclusive as a result of the incompleteness of both planet detection and stellar companion detection. Some studies suggested a significant deficit of stellar companions for the planet-hosting sample in contrast to the field population \citep[e.g.,][]{Roell2012, Wang2014a, Wang2014b, Kraus2016}, while others found no clear evidence of distinguishable planet frequency between binaries and single stars except for close stellar companions <100 au \citep[e.g.,][]{Bonavita2007, Horch2014, Asensio-Torres2018, Bonavita2020}. Synthesizing previous surveys of planet hosts, \citet{Moe2021} found planet suppression to be a gradual function of binary separation within $\sim$200 au and no influence from wide companions beyond 200 au. 
It is also suggested that the impact of stellar companions becomes more important for higher-mass planets since massive close-in giant planets (hot Jupiters) are more common in multiple systems \citep{Ngo2016, Fontanive2019, Ziegler2020}, implying that either giant planet formation is more efficient in wide binaries, or dynamical perturbations from stellar companions more likely result in orbital migration of giant planets \citep{Fontanive2019, Su2021}.
In terms of the distribution of planet properties, \citet{Desidera2007} and \citet{Ngo2017} indicated that planet parameters such as mass, period, and eccentricity are indistinguishable between single and multiple systems, while \citet{Moutou2017} suggested that the multiplicity of eccentric-planet hosts is higher than circular-planet hosts, as expected from numerical simulations \citep{Kaib2013}. Therefore the trend remains to be explored. 

Directly capturing the interactions between disks and stellar companions opens a window into the starting conditions of the planet formation process in these complex systems. Here we present the detection of circumstellar dust around three T Tauri multiple systems for the first time, using polarimetric differential imaging with SPHERE/IRDIS on ESO's Very Large Telescope (VLT) \citep{Beuzit2019} as part of the Disk Evolution Study Through Imaging of Nearby Young Stars (DESTINYS) programme \citep{Ginski2020, Ginski2021b}. 
In the DESTINYS samples, we did not specifically select targets based on their stellar multiplicity.
The three multiple systems are covered in the survey because they meet the criteria that the stars are optically bright members of nearby young star-forming regions and show near and mid-infrared excesses in their SED.
The snapshots of these systems provide an excellent comparison of the diverse morphology of disks in different evolutionary stages and binary configurations. Combined with analyses of age, mass, and orbital motion of the stellar systems, the structures of the disks provide further indications of the impact of stellar companions on the evolution of circumstellar disks and prospects for planet formation in such systems. 

This paper is organized as follows. In Section~\ref{sec:age_mass}, we introduce the properties of the three targeted multiple systems, namely, \CHX, \SCRA, and \HPCHA, with a particular focus on age and mass. The observations and data reduction are described in Section~\ref{sec:observations}. We present the analyses of polarimetric and astrometric data of each system in Section~\ref{sec:results}. These in conjunction with the knowledge of system properties are considered in Section~\ref{sec:discussion} to infer the disk structure and evolution history. Comparing the three systems, we finally discuss the implications for planet formation in binary systems. The summary is presented in Section~\ref{sec:conclusion}.

\begin{table*}[ht!]
\caption{Properties of the targeted young binary (or triple) systems.}
\label{tab:object}      
\centering 
\resizebox{\textwidth}{!}{
\begin{tabular}{c c c c c c c c c}     
\hline\hline                
Object & RA (h m s) & Dec (d m s) & Parallax (mas) &  Age (Myr)&  Separation (au) & SpT  &  Mass ($M_\odot$) & Ref. \\    % table heading 
\hline
\CHX &  11 12 42.69  & $-$77 22 23.1 & $5.2302 \pm 0.1675$ & $\sim$5  & $\sim$50  & G8 \& K7  &  1.8 \& 0.8 & 1, 4 \\
\SCRA &  19 01 08.60 & $-$36 57 19.0  & $ 6.2291 \pm 0.0683$ & $\sim$1  & $\sim$200 &  K6 \& K6  & 1 \& 1 & 2 \\
\HPCHA &  11 08 14.94 & $-$77 33 52.2 & $5.3381	\pm 0.0352$ & $\sim$10  & $\sim$500 & G5 \& M1 \& M3.5  &  1.4 \& 0.8 \& 0.3 & 3, 4 \\
\hline                                  
\end{tabular}
}
\tablebib{ Parallax: \citet{GaiaCollaboration2022}, (1) \citet{Daemgen2013}, (2) \citet{Gahm2018}, (3) \citet{Schmidt2013}, (4) \citet{Manara2017} }
\end{table*}

\section{Individual system properties}\label{sec:age_mass}

In order to determine the evolutionary stage of the targets, we summarized previous studies of the properties of each system, in particular, the age and stellar mass, as also listed in Table~\ref{tab:object}.

\subsubsection*{\CHX}

\CHX\ is a weakly (or non) accreting T Tauri star with a close binary companion separated by $\sim0.24$\arcsec\ ($\sim$50 au) \citep{Daemgen2013}.  
\CHX\ is located in the Chamaeleon I (Cha I) cloud, which is among the nearest young star-forming regions at a distance of $192\pm6$ pc \citep{Dzib2018}. The age distribution of Cha I was previously determined to peak around 3$-$4 Myr \citep{Luhman2007} based on an old distance estimate of $160\pm15$ pc \citep{Whittet1997}. The updated GAIA DR2 distance would result in a luminosity increase of $\Delta \log L \sim 0.15$, therefore implying a younger age.
Assuming a distance to Cha I of 160 pc, \citet{Daemgen2013} estimated the ages to be $16^{+30}_{-10}$ and $9^{+16}_{-6.5}$ Myr for the A and B components respectively, and the masses of 1.3 and 0.8 $M_\odot$ using \citet{Siess2000} isochrones. \citet{Manara2017} derived a mass of 1.62 $M_\odot$ for the A component using the \citet{Baraffe2015} isochrones. 
Taking into account the distance change, we scaled the mass and age using the luminosity and effective temperature reported the in the literature and \citet{Siess2000} isochrones. We adopted the masses of 1.8 and 0.8 $M_\odot$ for the binary and the age of $\sim$5 Myr.

\subsubsection*{\SCRA}

\SCRA\ is an archetypal system of the classical T Tauri stars with rich emission lines as a result of the accretion from circumstellar disks and emission from stellar or disk winds. 
The stars show significant spectroscopic variability because of the variable extinction from the circumstellar dust \citep{Gahm2008}. 
The binary separation is $\sim1.3$\arcsec\ ($\sim$200 au).
The stellar parameters have been extensively studied with resolved high-resolution spectra of both stars \citep{Gahm2018, Sullivan2019}. 
The study by \citet{Gahm2018} suggested that despite the different levels of veiling, both stars had quite similar properties with an effective temperature of 4250 K, and a mass of 1 $M_\odot$. 
\citet{Sullivan2019} estimated age of $\sim$1 Myr using the \citet{Baraffe2015} isochrones, determined different spectral types for the two stars (K7 \& M1), and obtained lower temperatures (4000 \& 3700 K) and masses (0.7 \& 0.45 $M_\odot$). 
We adopt the estimation by \citet{Gahm2018} since the total mass of $\sim2M_\odot$ is more consistent with our orbit analysis as presented in Section~\ref{sec:astrometry}.

\begin{table*}[h]
\caption{Observing log of three young binary (or triple) systems with SPHERE/IRDIS. }
\label{tab:obslog}      
\centering             
% \resizebox{0.5\textwidth}{!}{
\begin{tabular}{c c c c c c c c c}     
\hline\hline                
Object & Date & Filter & DIT (s) & NDIT & N$_\mathrm{cycle}$ & on-target time (min) & seeing (\arcsec) & $\tau_0$ (ms) \\    % table heading 
% & & & time (s) & &  &  & \\
\hline
\CHX & 2020-02-18 & BB\_H & 2 & 24  & 14  & 44.8  & 0.43 $-$ 0.96 & 11.5 \\
\CHX & 2020-02-28 & BB\_H & 8 & 8  & 13  & 55.5  & 0.6 $-$ 1.17 & 5.2 \\
\SCRA & 2021-09-08 & BB\_H & 8 & 2 & 42  & 44.8  & 0.46 $-$ 0.77 & 5.8\\
\HPCHA & 2020-01-13 & BB\_H & 64 & 1 & 14 & 59.7  & 0.33 $-$ 0.54 & 8.5 \\
\hline                                  
\end{tabular}
\tablefoot{DIT: integration time per exposure. NDIT: number of exposures per half-wave plate position. N$_\mathrm{cycle}$: number of full polarimetric cycles. $\tau_0$: atmospheric coherence time.}
% }
\end{table*}  

\subsubsection*{\HPCHA}

\HPCHA\ A, also named Glass I, is a G-type WTTS in the Cha I star-forming region, with a wide companion at 2.4\arcsec\ ($\sim$500 au) \citep{Chelli1988}. The companion itself was later resolved into a close binary B and C \citep{Daemgen2013, Schmidt2013}. The B and C components show significant variation in separation (from 70 to 180 mas), indicating a highly inclined and/or eccentric orbit. 
\citet{Schmidt2013} derived a mass of 1.26, 0.76, and 0.28 $M_\odot$ for the three stars and an age of $\sim$2 Myr. \citet{Manara2017} derived a mass of 1.26 $M_\odot$ for the A component and 1.00 $M_\odot$ for the BC component.
Similar to \CHX, we used the stellar temperature and luminosity reported in \citet{Manara2017} and scaled the mass and age according to the updated distance to Cha I cloud. We adopted a mass of 1.4, 0.8, and 0.3 $M_\odot$ respectively, and the age of $\sim$10 Myr. However, we caution that the age derived from evolutionary models is rather uncertain \citep{Pecaut2012}.

\begin{table}
\caption{Archival HST, VLT/NACO, VLT/SPHERE, and ALMA observations included in this work.}
\label{tab:obs_archival}      
\centering             
\resizebox{0.5\textwidth}{!}{
\begin{tabular}{c c c c c}     
\hline\hline                
Object & Epoch & Instrument & Program ID & Filter/Transition  \\    % table heading 
\hline 
\SCRA & 1999-02-21 & HST/WFPC2 & GO-7418 & F814W\\
\SCRA & 2002-10-29 & VLT/NACO & 70.D-0444(A) & $H$\\
\SCRA & 2010-08-08 & VLT/NACO & 085.C-0373(A) & $K_S$\\
\SCRA & 2012-06-09 & VLT/NACO & 089.C-0196(A) & $L$  \\
\SCRA & 2015-06-25 & VLT/SPHERE & 095.C-0787(A) & $K_S$ \\
\CHX & 2016-01-25 & ALMA & 2015.1.00333.S & CO $J=3-2$ \\
\HPCHA & 2015-05-21 & ALMA & 2013.1.01075.S & CO $J=3-2$ \\
\hline
\end{tabular}
}
\end{table}  
   
%--------------------------------------------------------------------
\section{Observations and data reduction}\label{sec:observations}

\subsection{SPHERE/IRDIS observations} \label{sec:obs_sphere}

We present new observations on three young binary (or triple) star systems, \CHX, \SCRA, and \HPCHA, with SPHERE/IRDIS in dual-beam polarimetric imaging mode in the $H$-band \citep{Dohlen2008, Langlois2014, deBoer2020, vanHolstein2020}. The observing setup was similar for the three targets.
The main observing sequence was performed in pupil-tracking mode, and the primary star behind a coronagraph with an inner working angle of 92.5 mas \citep{Carbillet2011, vanHolstein2017}. The science frames were followed by center calibration frames, flux calibration frames, and sky frames, taken without the coronagraph. 

We observed \CHX\ on 18 Feb 2020 with an integration time of 2s per exposure for a total of 24 exposures per half-wave-plate position. We obtained 14 full polarimetric cycles amounting to a total integration time of 44.8 min. The observations, however, were not deep enough to clearly reveal the circumstellar features. Therefore another epoch was taken on 28 Feb 2020, with an integration time of $8\times8$ s per half-wave-plate position, resulting in 13 polarimetric cycles during the total 55.5 min on-target time. The integration time of the flux reference frame was set to 4 s to prevent saturation. 
Similarly, \HPCHA\ and \SCRA\ were observed on 13 Jan 2020  and 8 Sep 2021, with an integration time of 64 s and $8\times2$ s, amounting to 14 and 42 full polarimetric cycles, and a total exposure time of 59.7 min and 44.8 min, respectively. 
The observations were performed under excellent weather conditions as summarized in Table~\ref{tab:obslog}. 

We reduced the data using the IRDIS Data Reduction for Accurate Polarimetry pipeline \citep[IRDAP,][]{vanHolstein2020} with the default setup. In principle, the data reduction for multiple stellar systems works similarly as for single systems. Whereas, one potential complication originates from the astrophysical stellar polarization. IRDAP removes the residual stellar polarization by directly measuring it in an annulus around the central star in the Stokes $Q$ and $U$ images. Yet this is not straightforward for multiple systems as the two (or multiple) components of the system may have different astrophysical polarization (for instance, due to different dust surroundings). This was indeed the case for \HPCHA, where we found that the degree of linear polarization measured around the primary star ($\sim 0.27\pm 0.1 \%$) significantly differed from that around the companions ($\sim 1.6\pm 0.2 \%$). Therefore, the \HPCHA\ data required two separate removals of stellar polarization centering on either the primary or the companions.
The calibration centered on the primary revealed the circumprimary disk, while the calibration centered on the companion better preserved details of the circumsecondary disk. After subtracting the polarized stellar halos around both components individually, we combined the two versions of stokes images weighted with a linear gradient as a function of the distance to its center of calibration. This allowed us to reveal the scattered light from the circumstellar material around both components simultaneously.

\subsection{Archival HST, VLT/NACO, and SPHERE data} \label{sec:obs_archive}

In order to extract the astrometric information and to constrain the potential orbital motion of the binaries, we exploited archival observations for a large span of epochs taken with VLT/NACO \citep{Lenzen2003, Rousset2003}, VLT/SPHERE, and Hubble Space Telescope (HST).
\SCRA\ was observed with the Wide Field and Planetary Camera (WFPC2) on the HST on 21 Feb 1999 \citep{Stapelfeldt1997}. 
In addition, we inspected the archival images of \SCRA\ taken with VLT/NACO on 29 Oct 2002, 8 Aug 2010, and 9 Jun 2012 in the $H$, $K_S$, and $L$ band respectively, and reduced them following the standard procedure (including sky subtraction, aligning of individual exposures, and stacking) using the ESO eclipse software package and the jitter routine \citep{Devillard1999}.  \SCRA\ has been previously observed with VLT/SPHERE on multiple epochs. We selected the one taken on 25 Jun 2015 in the $K$-band for the astrometric analysis to fill in the temporal gap between the NACO observations in 2012 and our new SPHERE observations in 2021. The archival observations are summarized in Table~\ref{tab:obs_archival}.

For \CHX\ and \HPCHA, we extracted the astrometric information from the published literature.
Both systems were observed with the 4-meter telescope at the Cerro Tololo Inter-American Observatory (CTIO) on 4 May 1994 \citep{Ghez1997}. \CHX\ was observed with VLT/NACO on 26 Mar 2006, 13 May 2007, 19 Feb 2008, and 20 Feb 2009 \citep{Lafreniere2008, Vogt2012}. \HPCHA\ was observed with VLT/NACO on 20 Feb 2003, 25 Mar 2006, 20 Feb 2008, and 20 Feb 2009 \citep{Correia2006, Lafreniere2008, Schmidt2013}. The astrometric measurements are compiled in Table~\ref{tab:astrometry}.

\subsection{Archival ALMA data} \label{sec:obs_alma}
We inspected archival ALMA observations of the three systems to obtain a better overview of the disk morphology. A brief summary of the observations is presented in Table~\ref{tab:obs_archival}.

\CHX\ was observed on 25 January 2016 and 30 March 2016 as part of program 2015.1.00333.S (P.I.: I. Pascucci). A non-detection of the millimeter continuum from this program was presented in \citet{Long2018}. We retrieved the raw data from the ALMA archive and reprocessed it with the ALMA pipeline using the CASA software package \citep{THECASATEAM2022}. We imaged the CO $J=3-2$ line at a velocity resolution of 0.25 km s$^{-1}$ with the \texttt{tclean} implementation of the multi-scale CLEAN algorithm \citep{Cornwell2008} and a robust value of 1.0. To handle the irregular emission pattern, we used CASA's auto-multithesh algorithm \citep{Kepley2020} to generate the CLEAN mask. After applying a primary beam correction to the resulting image cube, we generated a moment 0 map by summing up emission between LSRK velocities of 2.5 and 6.5 km s$^{-1}$, and a moment 1 map by calculating the intensity-weighted velocities. 

\HPCHA\ was observed on 21 May 2015 as part of the program 2013.1.01075.S (PI: Daemgen). The 880 $\mu$m continuum and $^{12}$CO $J=3-2$ line were observed in Band 7. The correlator was set up with four spectral windows (SPWs). The SPW covering the CO $J=3-2$ line was centered at 345.797 GHz and had a bandwidth of 468.750 MHz and channel widths of 122.07 kHz. The other three SPWs were centered at 334.016, 335.974, and 347.708 GHz, and each one had a bandwidth of 2 GHz and channel widths of 15.625 MHz. The cumulative on-source integration time was 1 minute.
We directly retrieved the data product from the archive and performed no further reprocessing, as the goal is not to present detailed analyses of the ALMA observations, but to show qualitative comparisons to scattered light images.

%--------------------------------------------------------------------
\section{Results} \label{sec:results}

\subsection{Polarimetric analysis} \label{sec:PI}

\begin{figure*}[ht]
   \centering
   \includegraphics[width=\hsize]{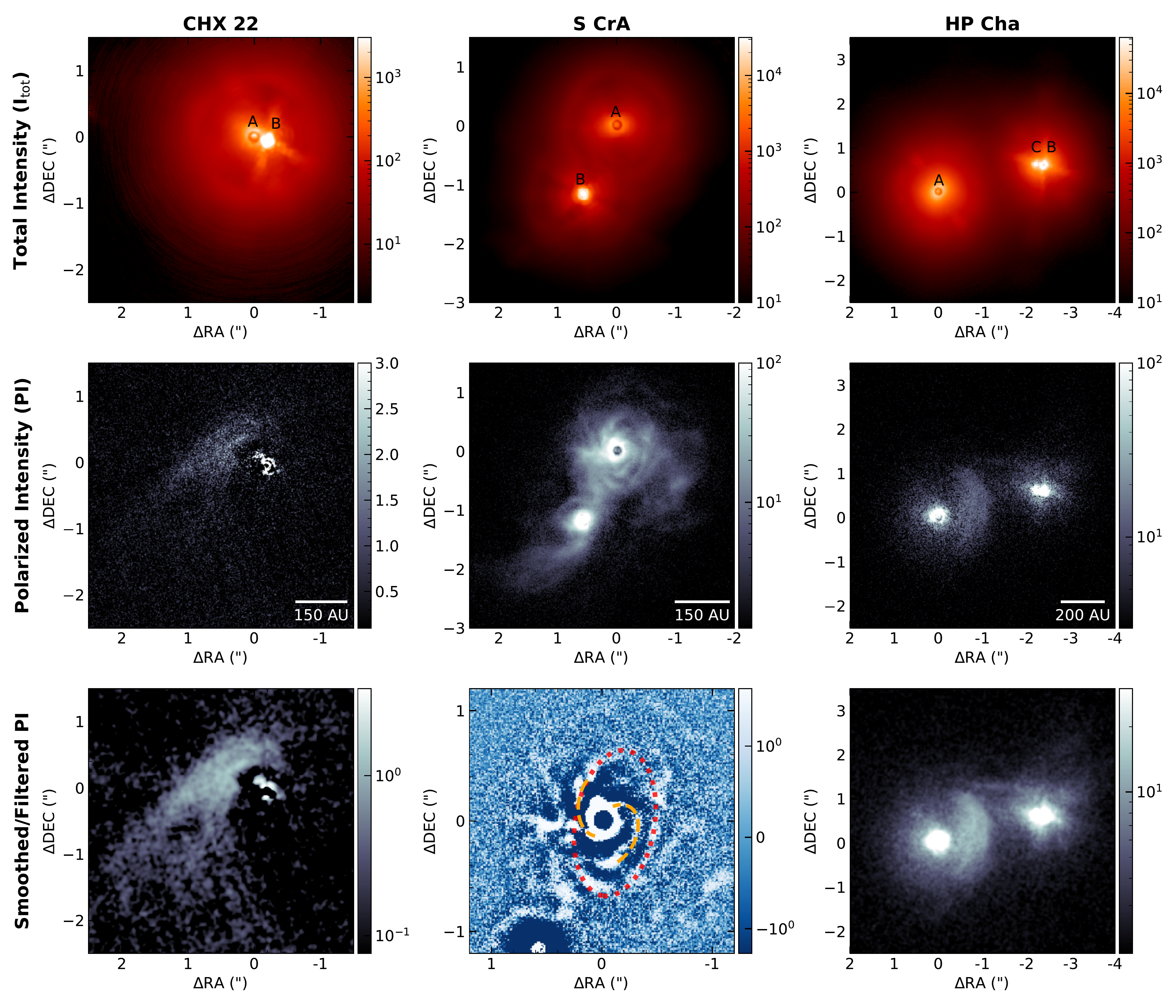}
      \caption{SPHERE/IRDIS images of the three multiple systems, centered on the primary stars. Each column shows one system, namely \CHX, \SCRA, and \HPCHA.
      Top row: total intensity images. Middle row: polarized intensity images.  Bottom row: processed PI images for better visualization of features. The PI image of \CHX\ is smoothed through convolution with a Gaussian kernel with a width of 2 pixels, revealing the dust tail clearly. The PI image of \SCRA\ is high-pass filtered by subtracting the Gaussian smoothed image from the original image, to highlight the potential ring (red dotted ellipse) and spiral features (orange dashed curves). The PI image of \HPCHA\ is also Gaussian smoothed. In the colorbars we show the scaling of the figures in arbitrary units to indicate the significance of the various detected structures. For a flux calibrated version of the polarimetric observations we refer to Appendix~\ref{app:stokes}. }
         \label{fig:polar_image}
   \end{figure*}

We detected polarized scattered light from the circumstellar disks around individual stars in the three systems. The total and polarized intensity images are presented in Fig.~\ref{fig:polar_image}.

\subsubsection*{\CHX}

\begin{figure}
   \centering
   \includegraphics[width=\hsize]{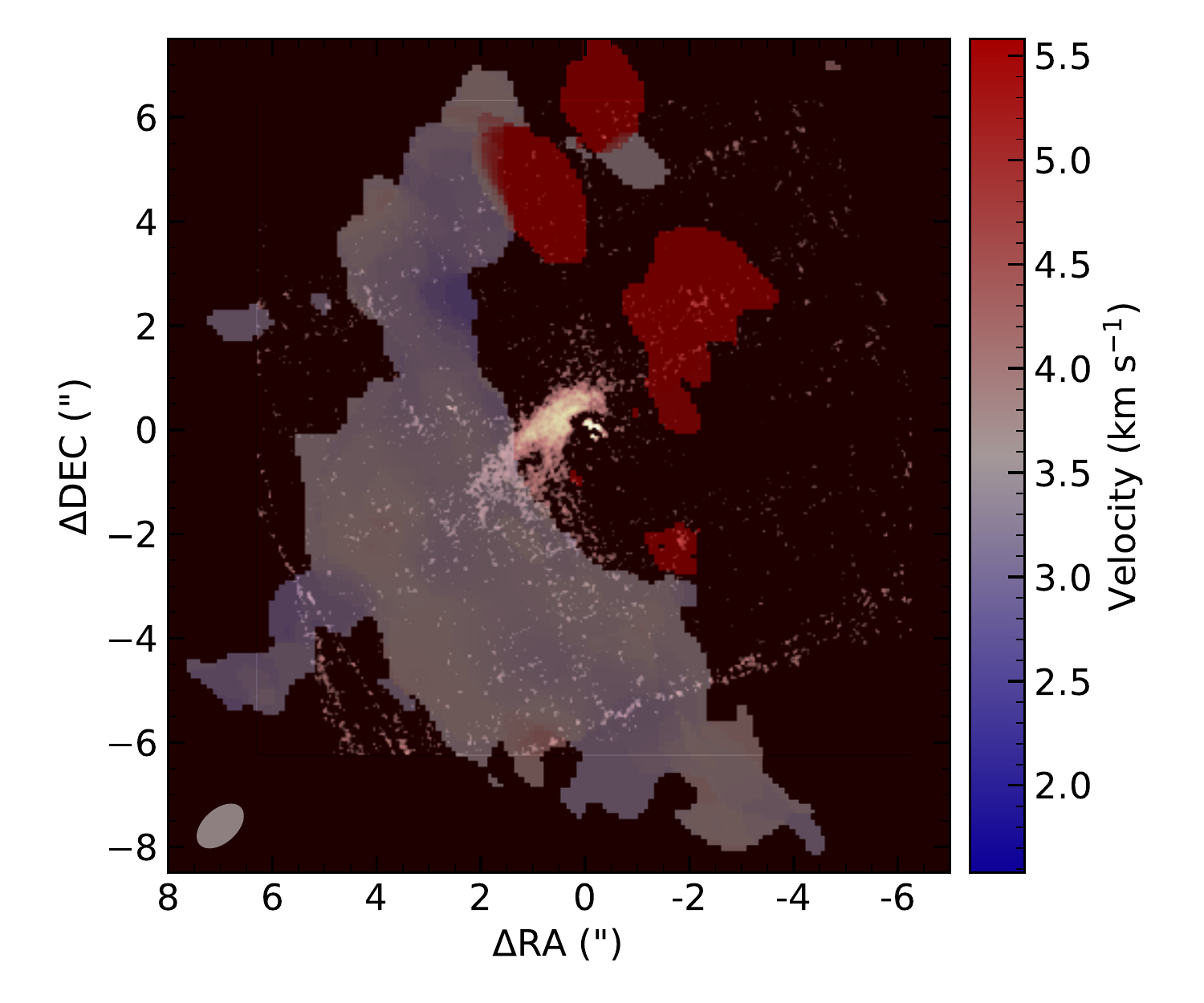}
      \caption{Overlay of SPHERE scattered-light image (background) and the ALMA moment 1 map of CO gas emission of \CHX\ (colormap centered on the systemic velocity of $\sim$3.6 \kms).
    %   The contours delineate the flux level of [3, 5, 10, 20]$\sigma$, where $\sigma$ is the standard deviation as measured in the image background. 
    The gray ellipse denotes the beam (1.07\arcsec $\times$ 0.65\arcsec) of ALMA observations.
    }
         \label{fig:chx_alma}
   \end{figure}

The circumstellar material forms a tail-like structure extending from the north to the southeast of the close binary. It is marginally resolved that the dust tail splits into two streams at the southern end as the scattered signal decays and merges into the noise.
The dust tail extends out by $\sim$650 au, which is larger than the scale of binary separation by an order of magnitude.
\CHX\ has also been observed in the mm wavelength with ALMA showing non-detection in the continuum \citep{Pascucci2016, Long2018}, while the extended emission from CO gas was detected. The overlay of the SPHERE image and the ALMA CO emission is shown in Fig.~\ref{fig:chx_alma}.

\subsubsection*{\SCRA}

The disturbed disks with a profusion of fractured structures are present in \SCRA. In the primary disk, we tentatively identified an elliptical ring (PA $\sim-9.3$ deg) as well as two spirals interior to that ring, as delineated in the high-pass filtered image as shown in the bottom-central panel of Fig.~\ref{fig:polar_image}. The ring was determined by fitting an ellipse to the polarized intensity (PI) image of the circumprimary disk. The center of the star is offset from the center of the ellipse along the minor axis, indicating that the ringed structure has a certain vertical height and is inclined with respect to the line of sight by $i\sim57$ deg. Utilizing the projected offset ($u$) of the ellipse-center from the star-center, we traced the height ($h$) of the elliptical ring via the relation $u=h\sin{i}$ \citep{deBoer2016}. It suggests that the disk height is $\sim$23 au at the radius of $\sim$107 au, meaning an aspect ratio $h/r \sim 0.21$, which is at the high end of T Tauri disks and similar to disks around IM Lup and MY Lup \citep{Avenhaus2018}. The spectral energy distribution (SED) of the system also indicates a flared disk \citep{Sicilia-Aguilar2013}.

The disk around the south-east stellar companion is connected with the material in the circumprimary disk, and features a streamer towards the south-east direction. The streamer is also marginally visible in the total intensity images of the system. We compared the HST/WFPC2 image taken in 1998 against the SPHERE/IRDIS observation in 2021, and found no obvious change in the shape over the decades, as shown in Fig.~\ref{fig:scra_hst}. However, we do notice the slight movement of the position angle of the companion through the offset between the peak signal and the overlaid contour. %The orbital motion is further investigated in Section~\ref{sec:astrometry}.

\subsubsection*{\HPCHA}

\begin{figure*}
   \centering
   \includegraphics[width=\hsize]{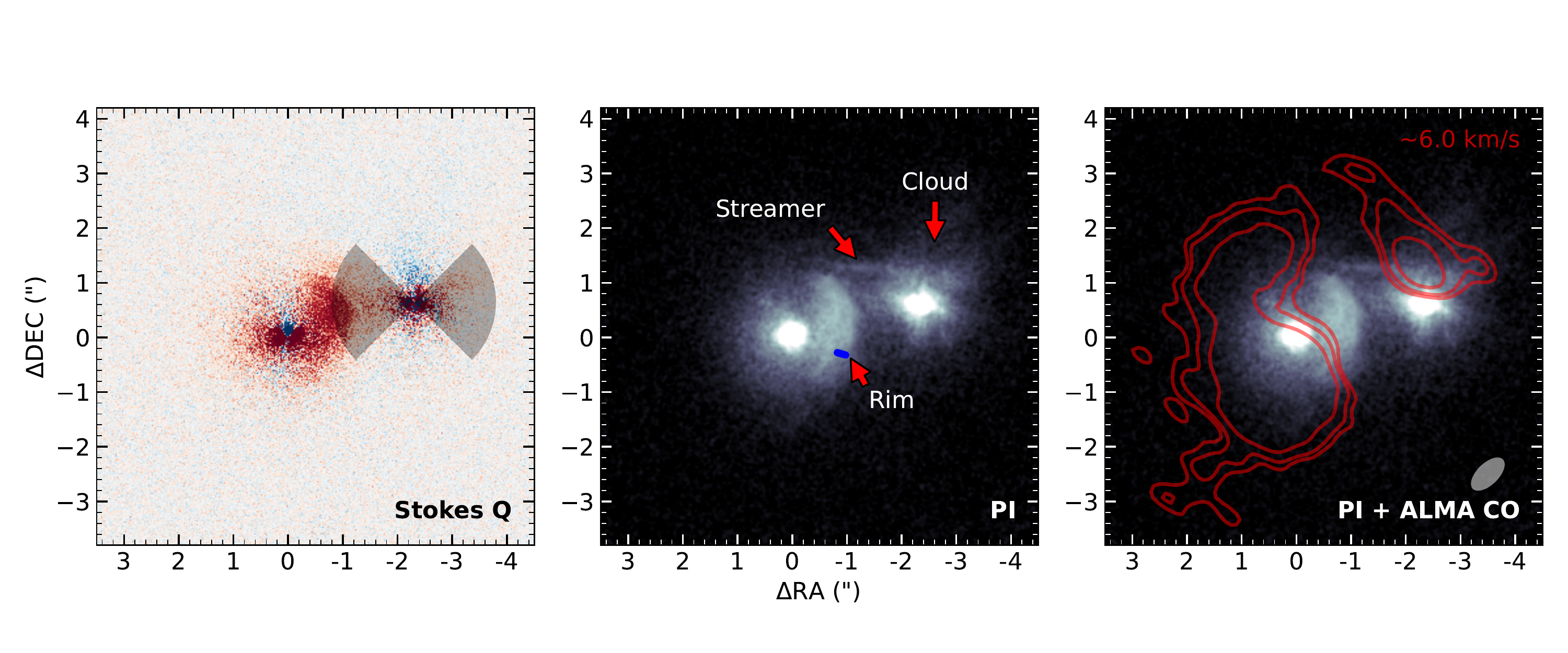}
      \caption{
      Left panel: Stokes $Q$ image of \HPCHA. The angle of linear polarization in the Stokes $Q$ pattern of the rim corresponds to an illumination source located at the position of companions BC (as delineated by the butterfly shadow centered at companions).
      Central panel: smoothed PI image where we identified the rim in the circumprimary disk, the streamer connecting the circumprimary and circumsecondary disk, and the nebulous envelope surrounding the BC components. The blue bar delineates the disk height at the most extended point of the outer rim.
      Right panel: overlay of ALMA CO gas emission (red contour) at the radial velocity of $\sim$6.0 \kms. The contours delineate the flux level of [3, 5, 10, 50]$\sigma$, where $\sigma$ is the standard deviation as measured in the image background. The gray ellipse denotes the beam (0.77\arcsec $\times$ 0.40\arcsec) of ALMA observations.}
         \label{fig:hpcha_alma}
   \end{figure*}

The circumstellar disk around the primary is clearly detected with a flared shape as indicated by the bright rim at the west side (see the annotation in Fig.~\ref{fig:hpcha_alma}). As the Stokes $Q$ and $U$ patterns of the rim correspond to an illumination source located at the position of companions B and C, therefore the bright rim delineates the edge and height of the circumprimary disk which is illuminated by the companions from the exterior.
Due to the external illumination of the primary disk by the binary companion, we can draw conclusions on the orbital geometry of the system. Since we noted a clear loss of scattered light signal in the intermediate region of the circumprimary disk (i.e. between the inner disk region and the externally illuminated rim), we conclude that the companions BC must be located slightly behind the primary star along our line of sight, such that the intermediate region of the primary disk surface is not illuminated by the companion, therefore not as bright as the outer rim.
Taking advantage of this unique back-illuminated feature, we traced roughly by eye the disk height to be $h = 42 \pm 10$ au at the most extended point of the outer rim (as annotated in Fig.~\ref{fig:hpcha_alma}), which has been corrected for the disk inclination ($i\sim37$ deg) derived from the mm-wavelength continuum observation \citep{Francis2020}. %  r=0.28, dr=0.04, PA=162, i=37
We caution that the height estimate has a large uncertainty subject to the exact position of the rim edge that one identified.
We then estimated the size of the disk as seen in scattered light along the major axis to be $r\sim260$ au by measuring the scale before the intensity drops below $3\sigma$ over the background. The flaring of the disk is therefore $h/r\sim0.16$. 
The disk around \HPCHA\ is comparable to other T Tauri disks as presented in \citet{Avenhaus2018}.

The disk around the BC components is less regularly shaped, forming a nebulous structure surrounding the close binary. We marginally identify the major axis of the nebulous envelope to be east-west oriented, which coincides with the orbital plane of the close binary. The nebulous envelope around the companions is connected to the disk around the primary through a tenuous streamer as annotated in Fig.~\ref{fig:hpcha_alma}. 

\subsection{Astrometry and orbit analysis}\label{sec:astrometry}

We carried out astrometric measurements on the available observations and utilized them to constrain the orbits of the stellar companions. 
As our SPHERE/IRDIS main observing sequences were conducted with the primary star masked by a coronagraph (see Section~\ref{sec:obs_sphere}), thus we used the flux reference frames without the coronagraph for more precise extraction of the astrometry. For the archival NACO data taken without a coronagraph (see Section~\ref{sec:obs_archive}), the stacked image in each epoch was used for the astrometric analysis.

We fit both the primary and secondary stars simultaneously using a model composed of two Gaussian profiles as implemented in the \texttt{astropy} model fitting package. The separation and position angle were then calculated from the positions of the Gaussian centers, with the astrometric calibration (including the pixel scale and the true north correction) for each instrument and epoch taken into account. The applied astrometric calibration can be found in Table~\ref{tab:calib}.
Our results together with previous astrometric measurements found in the literature are presented in Table~\ref{tab:astrometry}.

\begin{table}
\caption{Astrometry of the targeted multiple-star systems as extracted from our SPHERE/IRDIS observations as well as archival data taken with SPHERE, NACO, and HST.}
\label{tab:astrometry}      
\centering 
\resizebox{0.5\textwidth}{!}{
\begin{tabular}{l c c c c l}     
\hline\hline                
Object & Epoch (MJD) & Sep (mas) & PA (deg)  & Ref. \\   
\hline
\CHX & 49470   & 270 $\pm$ 30 & 236$\pm$  5 & 2  \\
& 53820.36777  & 246.9 $\pm$ 3.3 & 246.743 $\pm$ 1.245  & 2 \\
& 54233.98285  & 245.5 $\pm$ 3.6 & 247.312 $\pm$ 1.338  & 2 \\
& 54515.19528  & 244.0 $\pm$ 3.7 & 247.635 $\pm$ 1.401  & 2 \\
& 54882.37337  & 241.9 $\pm$ 3.9 & 247.991 $\pm$ 1.483  & 2 \\
& 58898.24507  & 212.6 $\pm$ 3.4 & 252.706 $\pm$ 0.944  & 1 \\
& 58908.16592  & 213.4 $\pm$ 2.0 & 252.506 $\pm$ 0.559  & 1 \\
\hline
\SCRA & 51230.96126  & 1329.1 $\pm$ 3.3 & 150.071 $\pm$ 0.142  & 1 \\
& 52576.05926  & 1322.8 $\pm$ 9.8 & 150.915 $\pm$ 0.417  & 1 \\
& 55416.21256  & 1312.6 $\pm$ 12.8 & 152.057$\pm$  0.571 & 1 \\
& 56087.38416  & 1313.9 $\pm$ 3.7 & 152.347 $\pm$ 0.155  & 1 \\
& 57198.31135  & 1316.8 $\pm$ 1.5 & 152.877 $\pm$ 0.112  & 1\textsuperscript{\textdagger} \\
& 57215.40732  & 1309.3 $\pm$ 1.3 & 153.021 $\pm$ 0.059  & 4 \\
& 59466.04295  & 1304.9 $\pm$ 1.3 & 153.979 $\pm$ 0.140  & 1 \\
\hline 
\HPCHA\ A(BC) & 49476   & 2500 $\pm$ 500 & 284$\pm$  5 & 3  \\
& 52690.20849 & 2430 $\pm$ 2 & 285.1  $\pm$ 0.5  & 3 \\
& 53819.41198 & 2421.5 $\pm$ 32.9 & 285.077 $\pm$  1.244 & 3  \\
& 54516.39621 & 2429.2 $\pm$ 36.9 & 285.103 $\pm$  1.401 & 3  \\
& 54882.40637 & 2425.7 $\pm$ 38.8 & 285.084 $\pm$  1.482 & 3  \\
& 58861.32325 & 2432.5 $\pm$ 3.3 & 284.417  $\pm$ 0.156  & 1 \\
\HPCHA\ B(C) & 53819.41198  & 70.2  $\pm$ 10.9 & 94.834 $\pm$  2.801 & 3 \\
& 54882.40637  & 98.4  $\pm$ 1.6 & 87.322  $\pm$ 1.497  & 3\\
& 58861.32325  & 179.2 $\pm$  4.4 & 83.841 $\pm$  1.374 & 1 \\
\hline
\end{tabular}}
\tablefoot{\textsuperscript{\textdagger} This epoch was excluded from the orbit fitting as it showed significant deviation from the well-constrained orbital motion (see Fig.~\ref{fig:orbit}). This is likely attributed to the inaccurate calibration of the pixel scale for SPHERE observations at the $K_S$-band.}
\tablebib{(1) this work, (2) \citet{Vogt2012}, (3) \citet{Schmidt2013}, (4) \citet{Sullivan2019}.}
\end{table}

\begin{table}
\caption{Calibration applied in our astrometric analysis.}
\label{tab:calib}      
\centering          
\resizebox{0.5\textwidth}{!}{
\begin{tabular}{c c c c c c}     
\hline\hline                
Epoch &  Instrument & Filter & Pixel scale & True north  & Ref. \\  
 &   &  & (mas\,pixel$^{-1}$) & correction (deg) &  \\ 
\hline
2002-10-29 & VLT/NACO & $H$ & $27.01 \pm 0.05$ & $0.08 \pm 0.18$ & 1 \\
2010-08-08 & VLT/NACO & $K_S$ & $13.231 \pm 0.020$ & $0.67 \pm 0.13$ & 2 \\
2012-06-09 & VLT/NACO & $L$ & $27.193 \pm 0.059$ & $0.568 \pm 0.115$ & 3 \\
2015-06-25 & VLT/SPHERE & $K_S$ & $12.267 \pm 0.009$ & $-1.75 \pm 0.10$ & 4 \\
\hline                                  
\end{tabular}
}
\tablebib{ (1) \citet{Chauvin2010}, (2) \citet{Ginski2014}, (3) \citet{Maire2016}, (4) \citet{Launhardt2020}
}
\end{table}  

With the astrometric measurements spanning across two decades, we utilized the \texttt{orbitize!} package \citep{Blunt2020} to constrain the orbital configuration of the binary systems.
We used the OFTI sampling method \citep{Blunt2017} with $10^5$ runs. The parallax and the prior system mass involved in the orbit fitting were adopted from Table~\ref{tab:object}. 
The orbital motion together with OFTI solutions, and the resulting posterior distributions of semi-major axis, inclination, and eccentricity for each individual system are shown in Fig.~\ref{fig:orbit} and \ref{fig:corner}.

\subsubsection*{\CHX}

As shown in Fig.~\ref{fig:corner}, the resulting posterior distributions suggest two groups of solutions for the eccentricity, one clustering around 0.4, and the other higher than 0.8. However the \texttt{orbitize!} tool is not well suited for eccentricities close to 1, and we suspect the accumulation of solutions towards the upper edge of the prior to be artificial. In order to better explore this issue, we fit the astrometry with another MCMC-based code developed by \citet{Beust2016}, that makes use of universal Keplerian variables instead of classical ones \citep{Danby1987}, and thus can naturally handle highly eccentric as well as unbound orbits with $e\geq1$ in a continuous manner. The posterior distributions are shown in Fig.~\ref{fig:corner_chx_unbound}. Now without the prior limitation $e<1$, the peak close to $e\sim1$ has disappeared, while the one at $e\sim0.4$ persists. The consistent peaks around $e\sim0.4$ with both codes suggests that the bound solutions with intermediate eccentricities are favored. 
Moreover, it is not surprising to find some solutions with unbound orbits due to the paucity of astrometric data and the small coverage of the orbit. However the fact that we note a sharp drop at $e=1$ in the posterior distribution of the eccentricity is a strong indication in favor of a bound orbit. 

\subsubsection*{\SCRA}

\begin{figure}[ht!]
  \centering
  \includegraphics[width=0.8\hsize]{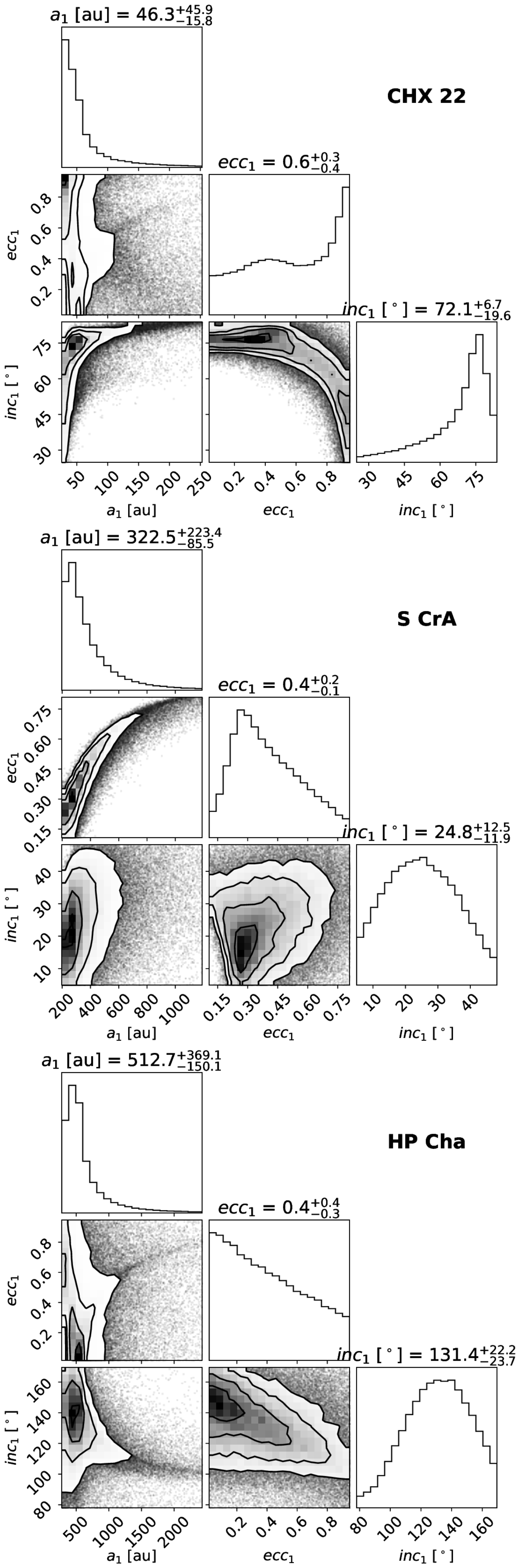}
      \caption{Posterior distributions of semi-major axis $a_1$, inclination $inc_1$, and eccentricity $ecc_1$ for the orbits of companions in \CHX, \SCRA, and \HPCHA.}
         \label{fig:corner}
  \end{figure}

The orbital separation and position angle of \SCRA\ show clear and steady motion over the baseline of two decades (see Fig.~\ref{fig:orbit}). The orbital solutions favor a small inclination $i=25\pm12$ deg and a moderate eccentricity $e=0.4^{+0.2}_{-0.1}$ as shown in Fig.~\ref{fig:corner}. 
The constraint on the system mass suggests a total mass of $\sim2 M_\odot$, which agrees with the estimation by \citet{Gahm2018}.

\subsubsection*{\HPCHA}

\begin{figure*}
   \centering
   \includegraphics[width=\hsize]{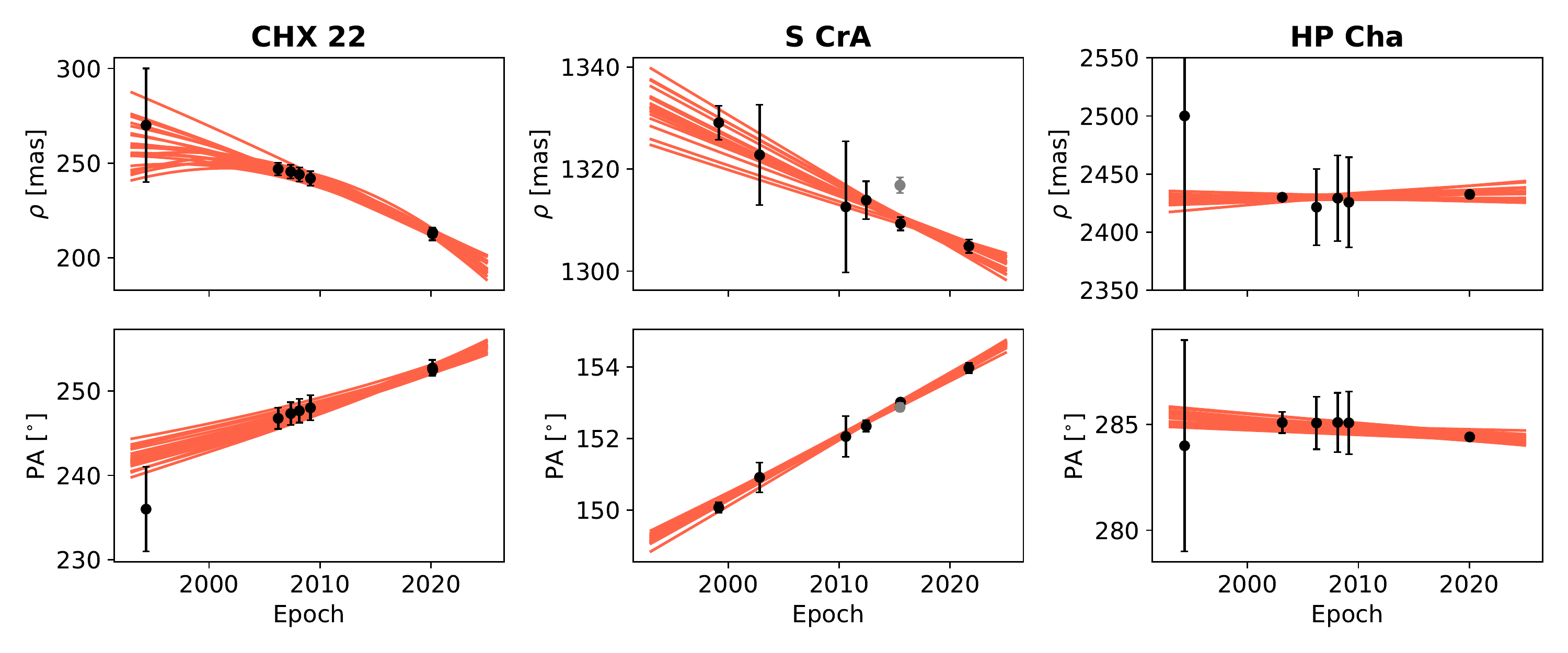}
      \caption{Separation (first row) and position angle (second row) of the companions relative to the primaries in \CHX, \SCRA, and \HPCHA\ versus time. The astrometric data were summarized in Table~\ref{tab:astrometry}. The red lines represent twenty randomly selected OFTI orbital solutions. The gray astrometric data points in \SCRA\ were excluded from the orbital fitting as explained in Table~\ref{tab:astrometry}}
         \label{fig:orbit}
   \end{figure*}

\begin{figure}%[ht!]
   \centering
   \includegraphics[width=\hsize]{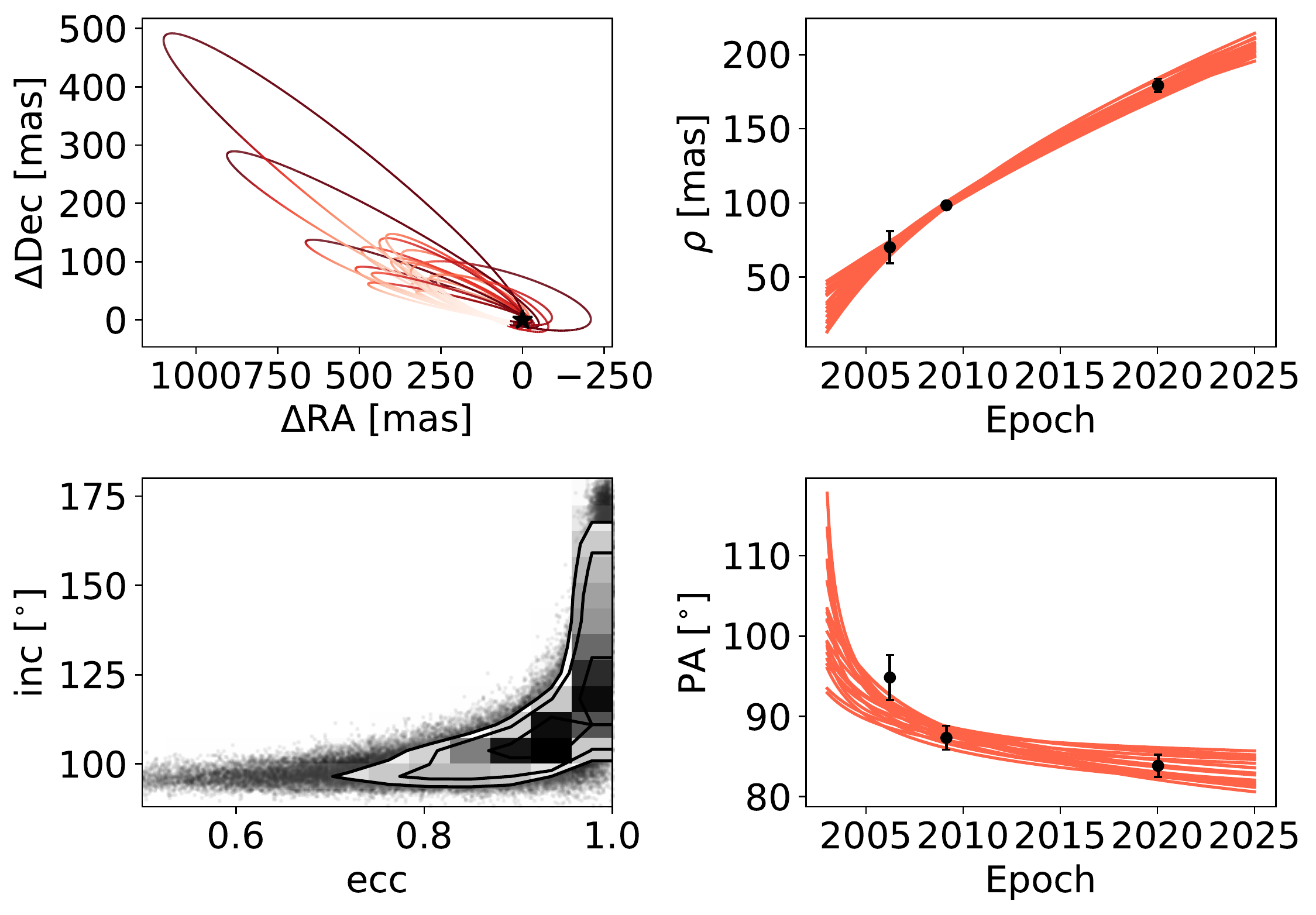}
      \caption{Orbital constraints of the binary \HPCHA\ BC. Top left, top right, and bottom right panel: the orbital motion in RA-Dec space, the change of separation, and position angle with time. The red lines represent twenty randomly selected OFTI solutions. Bottom left panel: the posterior distribution of inclination versus eccentricity.}
         \label{fig:orbit_hpcha_bc}
   \end{figure}

The astrometric measurements of the BC component with respect to the primary barely show any orbital motion beyond the uncertainties because of the large separation (see Fig.~\ref{fig:orbit}). Therefore, the orbital configuration around the primary is not well-constrained. However, the relative motion between B and C is apparent, suggesting a highly eccentric ($e\sim0.9$) and a nearly edge-on ($i\sim100^\circ$) orbit for the binary BC (see Fig.~\ref{fig:orbit_hpcha_bc}).

%--------------------------------------------------------------------
\section{Discussion} \label{sec:discussion}

\subsection{Individual systems}
\subsubsection*{\CHX}

The scale of the observed dust tail is drastically different from the scale of the inner binary system. It therefore raises the question of whether the dust is the tidal tail thrown out from the original circumstellar disk due to the interaction with the stellar companion, or if it is exterior material flowing into the binary system. Given the tail-like structure, system age, and orbital configuration, we discuss the following possible scenarios, 1) stellar flyby, 2) periodic perturbation, and 3) cloud capture. 

The first scenario refers to the case that the companion is in an unbound or extremely eccentric orbit. The circumstellar disk around the primary was tidally stripped during the parabolic or hyperbolic flyby event. For instance, simulations of prograde stellar flybys could result in tail-shaped structures for the ejected circumprimary material \citep{Clarke1993, Vorobyov2020}, resembling the observations of \CHX. Similar dust tails have been observed in young systems such as Z CMa \citep{Liu2016, Dong2022}, and are proposed to be induced by stellar flybys.
Assuming a typical open-orbit solution like shown in Fig.~\ref{fig:corner_chx_unbound}, we estimated that the temporal window of such a close encounter (<0.24\arcsec) is only around 50 years, in contrast to the age of the system ($\sim$5 Myr). The probability of capturing such a moment in time is rather small. 
Moreover, as we also noted in Section~\ref{sec:astrometry} that the binary orbit probably has an intermediate eccentricity, therefore we argue that the stellar flyby scenario is unlikely.
This leads to the second scenario that the close companion with an eccentric orbital motion may periodically strip away the disk material as it passes through the pericenter ($\sim$50 au, see Fig.~\ref{fig:corner_chx_unbound}). Again, given the age estimation of \CHX, it is less plausible that the relics of the continuously periodic interaction remain observable after such a long time. However, we acknowledge that the age derived from isochrones has large systematic uncertainties, hence we cannot rule out a younger age for the system. Nevertheless, this does not change the outcome that the primordial disk has likely gone, as suggested by the non-detection of the dust continuum emission \citep{Pascucci2016, Long2018}. A stringent dust mass upper limit of $\sim0.2M_\oplus$ was put, suggesting a deficit of large grains in the system. This adds to the implication that the circumprimary disk has already dispersed due to photoevaporation and/or been significantly truncated by the close companion \citep{Long2018}. It is intriguing that small grains still persist across an extensive scale. 

Both the above scenarios suppose that the dust tail is the outcome of the disk-binary interaction. Alternatively, we suggest that the circumstellar dust tail may be the infalling material from an encounter with nearby molecular clouds. The shape of the circumstellar material bending towards the stars remarkably matches the arc-shaped envelopes as simulated in \citet{Dullemond2019}. Such arc shapes have been observed in other systems including AB Aur \citep{Grady1999},  HD 100546 \citep{Ardila2007}, SU Aur \citep{Akiyama2019, Ginski2021b}, FU Ori \citep{Liu2016, Takami2018}, and DR Tau \citep{Mesa2022}.
According to the column density map of the Chamaeleon molecular cloud complex as observed with the Herschel space telescope \citep{AlvesdeOliveira2014}, \CHX\ resides in a moderately dust-rich region, therefore possibly undergoing such encounter with clouds. 
In ALMA CO emission of \CHX\ as shown in Fig.~\ref{fig:chx_alma}, we found large-scale gas emission associated to \CHX. The gas emission shares similar velocities as the systemic velocity of \CHX ($\sim$3.6 \kms). It forms a streamer-like structure from roughly North to South-West, confirming the presence of remnant cloud material in the direct vicinity of \CHX.

\subsubsection*{\SCRA}

The disk structure of \SCRA\ is full of intriguing details. In Section~\ref{sec:PI}, we identified potential ring and spiral features in the circumprimary disk. Here we attempt to assess the plausibility and discuss the implications provided that the features are real.
As discussed in Section~\ref{sec:PI}, the geometric center of the ring feature is offset from the stellar position along its minor axis, with the east side (left-hand side) of the ring closer to the stellar position. This not only implies a certain vertical height of the disk at the separation of the ring but also that the east side of the ring is the near side of the disk, namely, the side of the disk that is tilted toward the observer.
It makes sense intuitively that the east side (i.e. the near side) of the ring indeed shows higher brightness than the far side because of the stronger forward scattering. 
In order to quantify the brightness contrast due to the scattering phase functions, we measured the intensity within a small aperture (5 pixels) at both the near side and the far side of the ring, and compare the contrast ratio to that of several other disks with similar inclinations, such as HD 34282 \citep{deBoer2021}, LkCa15 \citep{Thalmann2014, Thalmann2016}, IM Lup \citep{Avenhaus2018}, and PDS 70 \citep{Hashimoto2012, Keppler2018}. The contrast ratio in \SCRA\ is $\sim$1.3, lower than that of $\sim$7 in HD 34282 and IM Lup, while similar to that of 1.6 in LkCa15 and 2.8 in PDS 70. Given the large scatter in different disks, it is inconclusive based on this argument whether the ellipse-like structure that we identified in \SCRA\ corresponds to a ring in the disk or not. 
However, while the brightness contrast between forward and back scattering is on the low end of disk observations, we note that indeed the side that should be forward scattering in this tentative ring (the east side) is the brighter side, as expected from dust scattering phase functions.
Combined with the offset along the minor axis, as expected from the viewing geometry of a ring in a flared disk surface, this gives a reasonable indication that we trace a ring in the disk around the primary component.
% SCrA 1.2456502
% PDS70 2.8211699
% IMLup 7.057747
% HD163296 5.3318787
% LkCa15 1.6057452
% HD34282 7.8204813

In addition, the two spiral structures also seem inclined, because the southern spiral appears foreshortened as launched from the nearside compared to the northern spiral. This appears similar to the simulated images of spiral features seen at such viewing angles as presented in \citet{Dong2016}. We performed deprojection of the PI image assuming a disk inclination of 57 deg (as inferred from the elliptical ring in Section \ref{sec:PI}) and a flaring index of 1.2 (a typical value as found in \citet{Avenhaus2018}) using the python package \texttt{diskmap} \citep{Stolker2016}. The deprojected image is presented in Fig.~\ref{fig:scra_deprojection}, where we estimated the pitch angle of the two spiral arms to be $\sim$30 deg. We caution that the pitch angle could change subject to the assumed disk parameters such as the actual inclination of the features. However, the uncertainty due to this inclination effect is marginal since the launching areas of the spirals are close to the major axis of the disk. 
Hydrodynamical simulations suggest that the pitch angle of spiral arms and the azimuthal separation between two arms positively correlate with the mass of the companion \citep{Zhu2015, Fung2015}.
The pitch angle of $\sim$30 deg and the nearly 180-degree separation between both arms are therefore consistent with the scenario that the spirals are launched by the perturbation from the stellar companion \citep{Fung2015}, as in the case of HD 100453 \citep{Benisty2017}.
 
If the inclined ring is real, it cannot be explained by the perturbation from the outer stellar companion. Instead, other mechanisms such as a second perturber are needed to induce the ringed structure \citep{Bae2022}. With our SPHERE observations reduced with the angular differential imaging (ADI) technique, we can rule out the presence of companions more massive than 8-10~$M_\mathrm{Jup}$ in between the A and B components, and $\sim$3-5~$M_\mathrm{Jup}$ beyond 300 au (see Sec~\ref{sec:ADI}). Whereas, this can be complicated by the fact that the companion may still be embedded in the disk at the young age of $\sim$1 Myr, which will significantly relax the mass limit. Therefore, further high-contrast imaging studies to search for planets in the system are interesting.

Through our orbit analysis of the \SCRA\ binary (see Section~\ref{sec:astrometry}), we found the orbital plane is likely close to face-on with an inclination of $25\pm12$ deg. 
This value is consistent with the inner disk inclination for both stellar components, $28\pm3$ deg and $22\pm6$ deg respectively \citep{GravityCollaboration2017}.
This is, however, in tension with the inclination ($\sim$53 deg) of the tentative ring and spiral features as seen in our SPHERE observations.
Interestingly, the spectroscopic analyses by \citet{Gahm2018} found the inclinations of the stars to be $\sim$60 deg, similar to that of the outer disk observed in the scattered light but misaligned with the inner disk observed in the NIR-interferometry.
The dust continuum emission from outer disks around both stars was observed with ALMA but unresolved \citep{Cazzoletti2019}, therefore providing no information on the disk inclination. The ALMA $^{12}$CO and $^{13}$CO emission is dominated by the signal from the extensive envelope, showing intricate kinematic patterns with streamers and outflows. Deeper observations via optically thin tracers such as C$^{18}$O are required to probe the embedded disks. A detailed analysis of the gas emission can be found in the work by Gupta et al. (submitted). In addition, searching for shock tracers such as SO and SO$_2$ \citep{Tafalla2010, Booth2018, Garufi2022} may be interesting as the streamer across the stellar companion appears to connect onto east side of the primary disk. Alternatively, future long-baseline ALMA observations resolving the outer disk in \SCRA\ and determining the disk inclination will help unravel the whole picture. 

If the spin-orbit misalignment is real, which is not uncommon in wide binary systems \citep{Hale1994, Justesen2020}, the circumstellar disks were initially aligned with the stellar rotation axis as the material collapsed around each star, and later got aligned with the orbital plane as a result of the gravitational torque by the companion \citep{Lubow2000}.
The misalignment of individual circumstellar disks with the binary orbital plane seems not so surprising \citep[e.g.,][]{Williams2014, Jensen2014, Brinch2016, Fernandez-Lopez2017, Manara2019, Rota2022}. Theoretical studies by \citet{Foucart2014} suggest that the large disk-orbit misalignment could be maintained over the entire disk lifetime. Therefore, circumstellar disks in wide binaries are not necessarily coplanar with the binary orbit despite the tendency towards alignment \citep{Jensen2020}. 

The appearance of the disk in \SCRA\ is reminiscent of GW Ori with misaligned rings as disturbed by gravitational interactions in the multiple star system \citep{Kraus2020}. In GW Ori, the ringed structures are attributed to the tearing caused by the misaligned orbital plane with the disk. Likewise, the stellar companion in \SCRA\ may also play an important role in warping the circumprimary disk. Dedicated hydrodynamic simulations of the system will be essential for understanding the delicate substructures and the influence of the binary-disk interaction.

% The accretion rate of the primary star is 7.8e-8, 1.28e-7 M_solar/yr, measured from Pa_beta and Br_gamma line respectively (on average 1e-7 M_solar/yr). For the secondary, it is 3.5e-8 M_solar/yr (Sullivan+2019). Dust masses around the primary and secondary are 102 and 95 $M_\oplus$  \citep{Cazzoletti2019} . The age is ~1Myr.  Assuming a gas-to-dust ratio of 100, the timescale for consuming up the disk is 3e5 yr for the primary.

\subsubsection*{\HPCHA}

The circumprimary disk around \HPCHA\ appears to be a transition disk with a single ring in the mm dust continuum \citep{Long2018, Francis2020}, while no dust continuum emission was detected around the BC components. We inspected the archival CO emission data observed with ALMA and found gaseous material around both the primary and the secondary components, as also revealed in our scattered light image. 
In Fig.~\ref{fig:hpcha_alma}, we overlaid the average image of four frequency channels from 345.7894 GHz to 345.7887 GHz, corresponding to the radial velocity of 5.72$-$6.32 \kms. 
The maps of more channels can be found in Fig.~\ref{fig:hpcha_alma_channel}. 
The gas emission also delineates a streamer connecting the primary and companions, while it extends further north and appears more curved than the streamer observed in scatter light. The slight offset may be because only the bottom side of the streamer structure is illuminated by the companions which are located behind the primary star along our line of sight.
Since the systemic velocity of \HPCHA\ is $\sim$3.8 \kms, therefore the gas emission shown in Fig.~\ref{fig:hpcha_alma} is red-shifted with respect to the systemic velocity, meaning the material in the streamer flowing from the circumprimary disk to the secondary binary (combined with our previous constraint from the scattered light that the binary is spatially located behind the primary).
The feeding flow from the primary disk may be responsible for the disturbed, nebulous envelope around the BC components. In addition, the extremely eccentric orbit of the close binary likely contributed to stirring up the material.  
On the other hand, the circumprimary disk maintains a regular shape, indicating that the interaction is not as violent as in the case of \SCRA.

\subsection{Binary-disk interactions and planet formation}

Following the sequence from \CHX\ to \SCRA\ and \HPCHA, we note that as the binary separation increases, the binary-disk interaction is less violent and destructive. The primitive disk in the close binary \CHX\ ($a<50$ au) has likely been truncated significantly and has dissipated via internal photoevaporation as there is no evidence for millimeter continuum and little NIR excess \citep[see the discussion in][]{Long2018}.  

\SCRA\ may be undergoing a similar truncating process that the circumprimary disk is losing material through the streamer connected via the companion. However, as the binary separation in \SCRA\ ($a \sim 300$ au) is much larger than that in \CHX, substantial material may still remain for a longer timescale. 
The companion truncates the circumprimary disk in a few orbital timescales ($\sim$5000 yr) as a result of Lindblad resonances \citep{Artymowicz1994}. We roughly estimated the expected disk size based on the analytical models by \citet{Artymowicz1994} and \citet{Manara2019}. Adopting the system parameters including equal stellar masses, $a\sim320$ au, and $e\sim0.4$, as derived from our orbital analysis in Section~\ref{sec:astrometry}, the outer edge of the primary disk would be truncated at around $0.16a - 0.21a$ ($50-70$ au) depending on the disk viscosity. As the current disk size is well above 100 au, the truncation is likely ongoing, and the material streams towards the circumsecondary disk, which can either end up around the companion or get stripped away through the south-east tail.

The mass transfer from the circumprimary disk to the companion is also observed in \HPCHA, with a more tenuous streamer than the one in \SCRA. The turbulent scene in \SCRA\ is plausibly attributed to the moderate orbital eccentricity $\sim$0.4.
The larger separation of the \HPCHA\ BC components to the primary ($a \sim 500$ au) may also contribute, such that the secular timescale as a result of disk truncation is different from that of \SCRA. 
The analytic prediction of the tidally truncated disk size for \HPCHA\ is pending due to the unconstrained orbital configuration. A conservative estimation implies a truncated circumprimary disk size of $0.35a\sim$175 au, which is smaller than the extent of the disk as observed in scattered light ($\sim$260 au).
The tidal truncation likely takes effects on shaping the primary disk as we note the sharp drop-off of flux at the edge of the rim (Fig.~\ref{fig:hpcha_alma}). Nevertheless, the circumprimary disk evolution in \HPCHA\ is probably not significantly altered by the stellar companions in contrast to \SCRA.

This decreasing influence on circumprimary disk evolution at larger binary separations agrees well with the statistics from millimeter-wavelength studies.
Surveys of protoplanetary disks in star-forming regions with ALMA have found that dust disks in multiple systems with separations less than 200 au tend to be more compact and less massive than the counterparts around single stars, while the influence is marginal in wide binary systems \citep{Cox2017, Akeson2019, Manara2019}.
The sizes of gas disks in multiple systems are consistent with predictions from tidal truncation by companions \citep{Rota2022}.
Studies by \citet{Rosotti2018, Zagaria2021a, Zagaria2021b} suggested that disks perturbed by close companions undergo faster gas and dust evolution than those around single stars. Therefore the disk lifetime may not be long enough to form planets in close binaries. 
%Furthermore, \citet{Zurlo2020, Zurlo2021} suggested that the effect from stellar companions is mostly evident for large and massive disks.

As a consequence of the various extents of binary-disk interactions, the planet formation in stellar multiple systems likely has diverse outcomes. The formation and survival of planets in close binaries are expectedly rare, subject to the tidal truncation of disks and the later excitation of high eccentricities. 
Although there are systems like \CHX\ with late-stage infall that may potentially form secondary circumbinary disks, they are probably far from massive enough for efficient planet formation. The stellar multiplicity rate of planet-hosting stars is suggested to be lower than field stars for close companion separations (<50 au), providing evidence of suppressed planet formation in systems with close-in companions \citep{Wang2014a, Wang2014b, Kraus2016}. The statistics are consistent with the expectation from disk observations. 

For binary systems with wide separations, the opportunity for planet formation is not significantly undermined. Statistical analyses tend to find no evidence for distinguishable planet occurrence or distribution of planet properties between single and binary systems \citep{Bonavita2007, Horch2014, Ngo2017}. This agrees well with our observations of non-disturbed circumprimary disks, for instance, \HPCHA, where the planet formation is expected to be not different from that around single stars.
On the other hand, there is growing evidence that giant planet formation may be boosted in wide binaries \citep{Ngo2016, Fontanive2019}. Studies on planetesimal dynamics in highly inclined ($i>30$ deg) binary systems suggest that despite the truncation of disks, the planetesimals could pile up within the stability radius to allow for planet formation \citep{Xie2011, Batygin2011}.
Moreover, the presence of stellar companions or planets may induce substructures that assist to concentrate dust particles, resulting in efficient streaming instability and formation of planetary embryos \citep{Johansen2007, Baehr2022}. 
We argue that the formation of a planetary object in such a disturbed disk is indeed likely given the ring-like feature we identified in \SCRA\ which may be caused by an unseen embedded object.
The stellar companion could also play an important role in the dynamical evolution of planetary embryos after the dispersion of the gas disk. The further growth of embryos to terrestrial planets could be enhanced in binary systems with intermediate separation \citep{Zhang2018}. Consequently, planet formation may even be enhanced in favorable circumstances. Future observations of newborn planets in binary systems such as \SCRA\ and \HPCHA\ will provide insights into the processes.

%--------------------------------------------------------------------
\section{Conclusion} \label{sec:conclusion}

We presented the detection of circumstellar dust in three multiple systems, \CHX, \SCRA, and \HPCHA, using polarimetric differential imaging with SPHERE/IRDIS. Combining the disk morphology with constraints on the age and binary orbits, we unraveled the evolution history and the impact of the stellar companions on the disks. 

\begin{itemize}

\item The dust in \CHX\ forms an extended tail-like structure surrounding the close binary. Considering the likely bound orbit of the companion and the age of the system, we suggest that the dust tail may be secondary material resulting from a late-stage cloud capture event, instead of being relics from the stellar flyby or tidal truncation. The primordial disk in the binary has probably dispersed or been significantly truncated via the interaction with the close companion. 

\item The violent interaction between disk and binary is well captured in \SCRA\ showing intricate disk structures as a consequence of perturbations by the companion with a moderately eccentric orbit. 
We presented evidence for a ring and spiral features in the disk. While the spirals are consistent with perturbations from the known stellar companion, the ring-like feature would require other mechanisms such as a second (lower mass) perturber in the system. 
The circumsecondary disk is truncating the primary disk, and material is transferred via a streamer connecting both disks. 
    
\item In contrast, \HPCHA\ with a larger binary separation shows a less fierce interaction. We detected a regularly shaped disk around the primary with a tenuous streamer feeding material towards the nebulous disk around the companions. 

\end{itemize}

The comparison of the three systems (spanning a wide range of binary separation from 50 to 500 au) illustrates the decreasing influence on disk structure and evolution with the distance of companions. It implies that planet formation in disks is likely obstructed around close binary systems, while barely undermined in wide binaries, which agrees with the statistical analyses of disk census and exoplanet population in binaries. 

%--------------------------------------------------------------------
\begin{acknowledgements}
    This paper makes use of the following ALMA data: ADS/JAO.ALMA\#2013.1.01075.S, ADS/JAO.ALMA\#2015.1.00333.S. ALMA is a partnership of ESO (representing its member states), NSF (USA) and NINS (Japan), together with NRC (Canada), MOST and ASIAA (Taiwan), and KASI (Republic of Korea), in cooperation with the Republic of Chile. The Joint ALMA Observatory is operated by ESO, AUI/NRAO and NAOJ.
    Y.Z. acknowledges funding from the European Research Council (ERC) under the European Union's Horizon 2020 research and innovation program under grant agreement No. 694513.
    Support for J. H. was provided by NASA through the NASA Hubble Fellowship grant \#HST-HF2-51460.001-A awarded by the Space Telescope Science Institute, which is operated by the Association of Universities for Research in Astronomy, Inc., for NASA, under contract NAS5-26555.
    A.Z. acknowledges support from the FONDECYT Iniciaci\'on en investigaci\'on project number 11190837 and ANID – Millennium Science Initiative Program – Center Code NCN2021\_080. 
    A.R. has been supported by the UK Science and Technology research Council (STFC) via the consolidated grant ST/S000623/1 and by the European Union's Horizon 2020 research and innovation programme under the Marie Sklodowska-Curie grant agreement No. 823823 (RISE DUSTBUSTERS project).
    M.B. acknowledges funding from the European Research Council (ERC) under the European Union’s Horizon 2020 research and innovation programme (PROTOPLANETS, grant agreement No. 101002188).
    CHR acknowledges the support of the Deutsche Forschungsgemeinschaft (DFG, German Research Foundation) Research Unit ``Transition discs'' - 325594231. CHR is grateful for support from the Max Planck Society.
    G.R. acknowledges support from the Netherlands Organisation for Scientific Research (NWO, program number 016.Veni.192.233). This project has received funding from the European Research Council (ERC) under the European Union's Horizon Europe Research \& Innovation Programme under grant agreement No 101039651 (DiscEvol).
    
\end{acknowledgements}

\bibliographystyle{aa} % style aa.bst
\bibliography{ref} % your references 

\appendix

\section{Morphology of disks in \SCRA}

   \begin{figure}[h]
   \centering
   \includegraphics[width=\hsize]{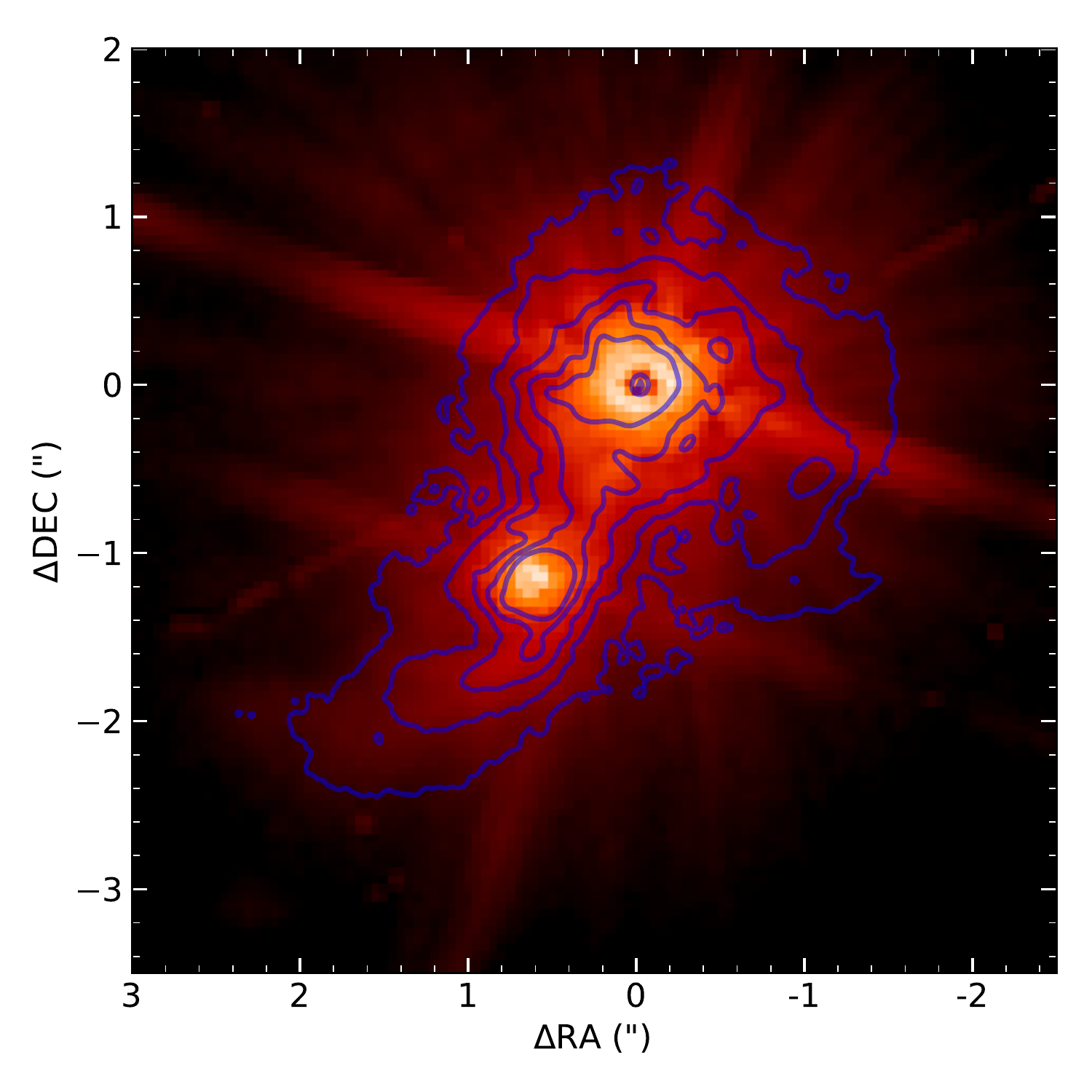}
      \caption{HST observations of \SCRA. The blue contour delineates the flux level of [5, 10, 20, 40, 80]$\sigma$ in the SPHERE PI image, where $\sigma$ is the standard deviation as measured in the image background.}
         \label{fig:scra_hst}
   \end{figure}

  \begin{figure}[h]
   \centering
   \includegraphics[width=\hsize]{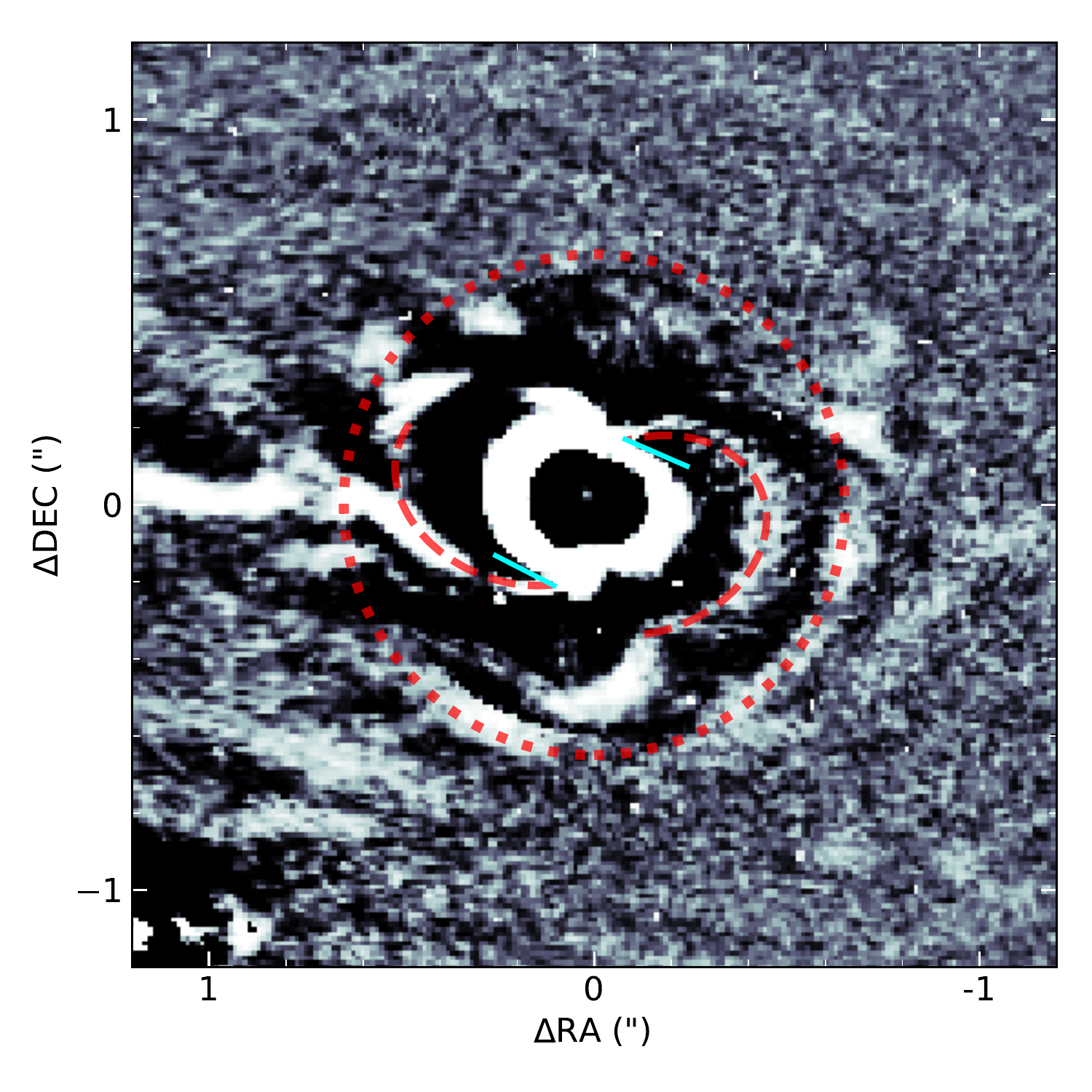}
      \caption{Deprojected PI image of \SCRA\ assuming a disk inclination of 57 deg. The ring and sprirals are delineated in red dashed lines. The pitch angle of spirals is estimated to be $\sim$30 deg.}
         \label{fig:scra_deprojection}
   \end{figure}

   \begin{figure}[h]
   \centering
   \includegraphics[width=\hsize]{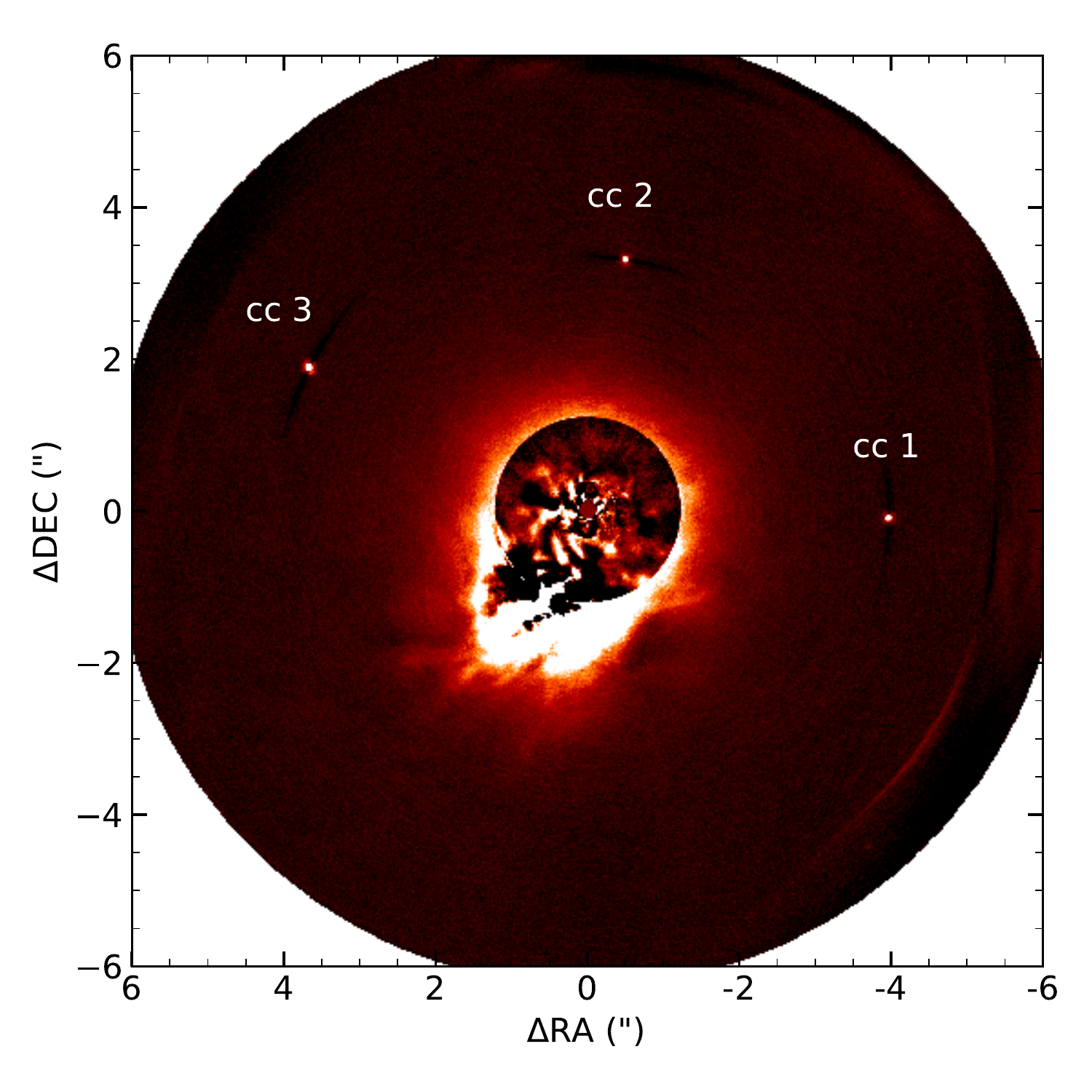}
      \caption{PCA image of \SCRA\ with two principal components subtracted.}
         \label{fig:scra_pca}
   \end{figure}
   
\section{SPHERE ADI imaging of the system \SCRA}\label{sec:ADI}

We performed a data reduction for the ADI images of the system around \SCRA. To process the data we used a custom routine to perform background subtraction, flat-fielding, centering of the images, and the VIP code \citep{GomezGonzalez2017} to perform the principal component analysis (PCA) to get the final images and contrast. 
The 5$\sigma$ contrast was calculated with the VIP code as the $\sigma \times$noise of the ADI image divided for the throughput. The calculation takes into account the small sample statistics correction presented in \citet{Mawet2014}. 

In the final processed image the stellar binary companion is clearly visible in the FoV, very saturated in the images. Also, three very likely background companion candidates (cc) are visible in the FoV (see Fig.~\ref{fig:scra_pca}). We report in Table~\ref{tab:astro} their coordinates. To calculate the astrometry we applied the same routines presented in \citet{Zurlo2020, Zurlo2021}. Further high-contrast imaging follow-up of the system will confirm the nature of the stars identified. 

\begin{figure}[h]
   \centering
   \includegraphics[width=\hsize]{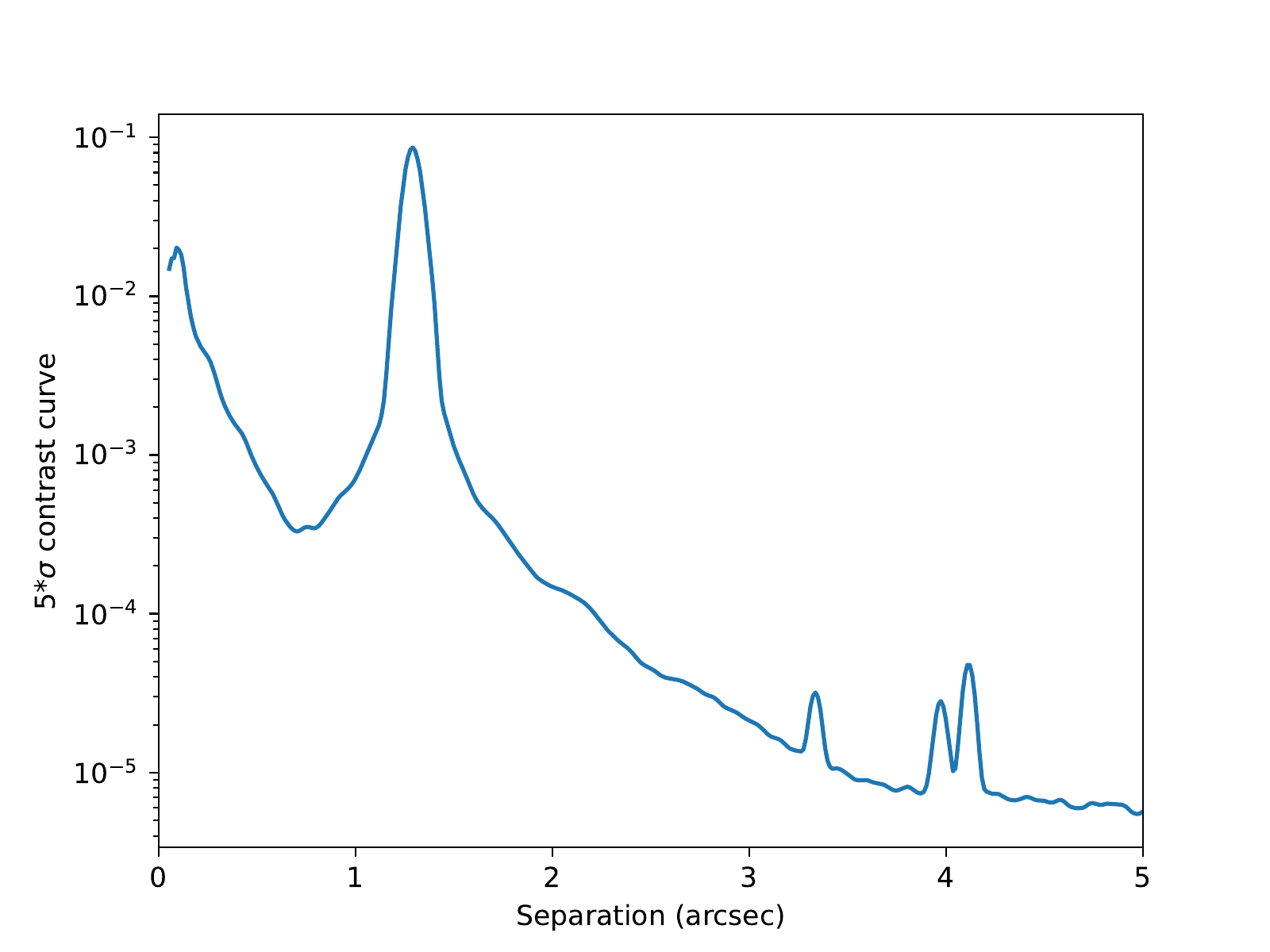}
      \caption{Contrast curve at 5$\sigma$-level for the ADI image of the system S CrA.}
         \label{fig:contr}
   \end{figure}

To calculate the mass limits for putative sub-stellar companions in the system, we used the contrast curves calculated for the PCA image, shown in Fig.~\ref{fig:contr} and used the COND evolutionary models for substellar companions as presented in \cite{Baraffe2003, Baraffe2015}. The age for the system used in the analysis is 1 Myr.

\begin{table}
\caption{Astrometric positions for the 3 (putative background) companion candidates identified in the ADI images.}
\label{tab:astro}      
\centering          
\begin{tabular}{c c c }     
\hline\hline                
Object  &  $\Delta$RA & $\Delta$Dec \\  
 &   (mas) & (mas)  \\ 
\hline
cc 1 & -4010 $\pm$ 5 & -172 $\pm$ 5 \\
cc 2 & -544 $\pm$ 5  &  3232$\pm$ 5   \\
cc 3 & 3632 $\pm$ 5 &  1829 $\pm$ 5 \\
\hline                                  
\end{tabular}
\end{table}  

   \begin{figure}[h]
   \centering
   \includegraphics[width=\hsize]{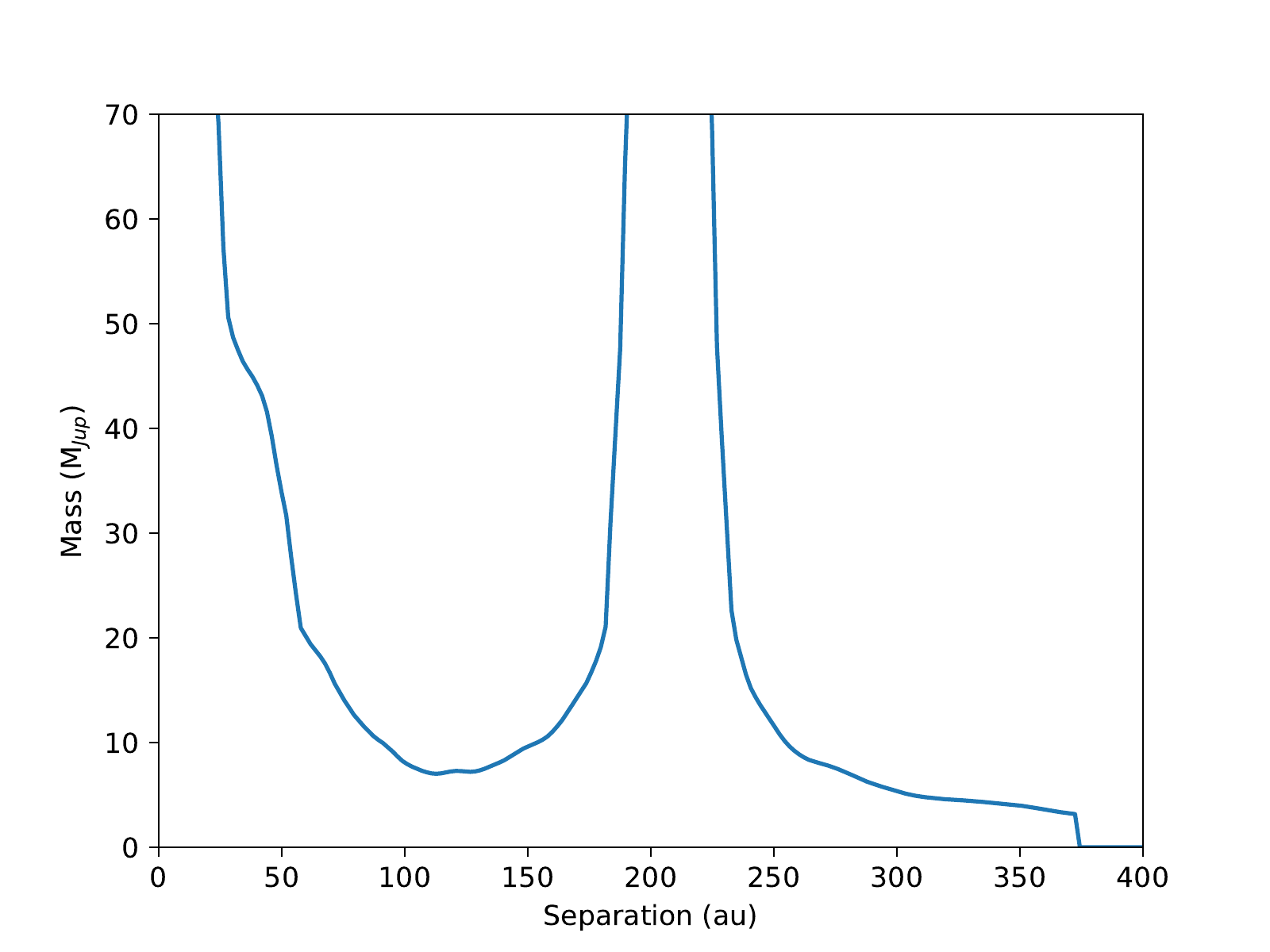}
      \caption{Mass limits for companions around the star S CrA, calculated from the contrast curve in Fig.~\ref{fig:contr}.}
         \label{fig:mass_contr}
   \end{figure}

\section{Orbit fitting for \CHX\ with the possibility of unbound orbits}

\begin{figure*}
   \centering
   \includegraphics[width=\hsize]{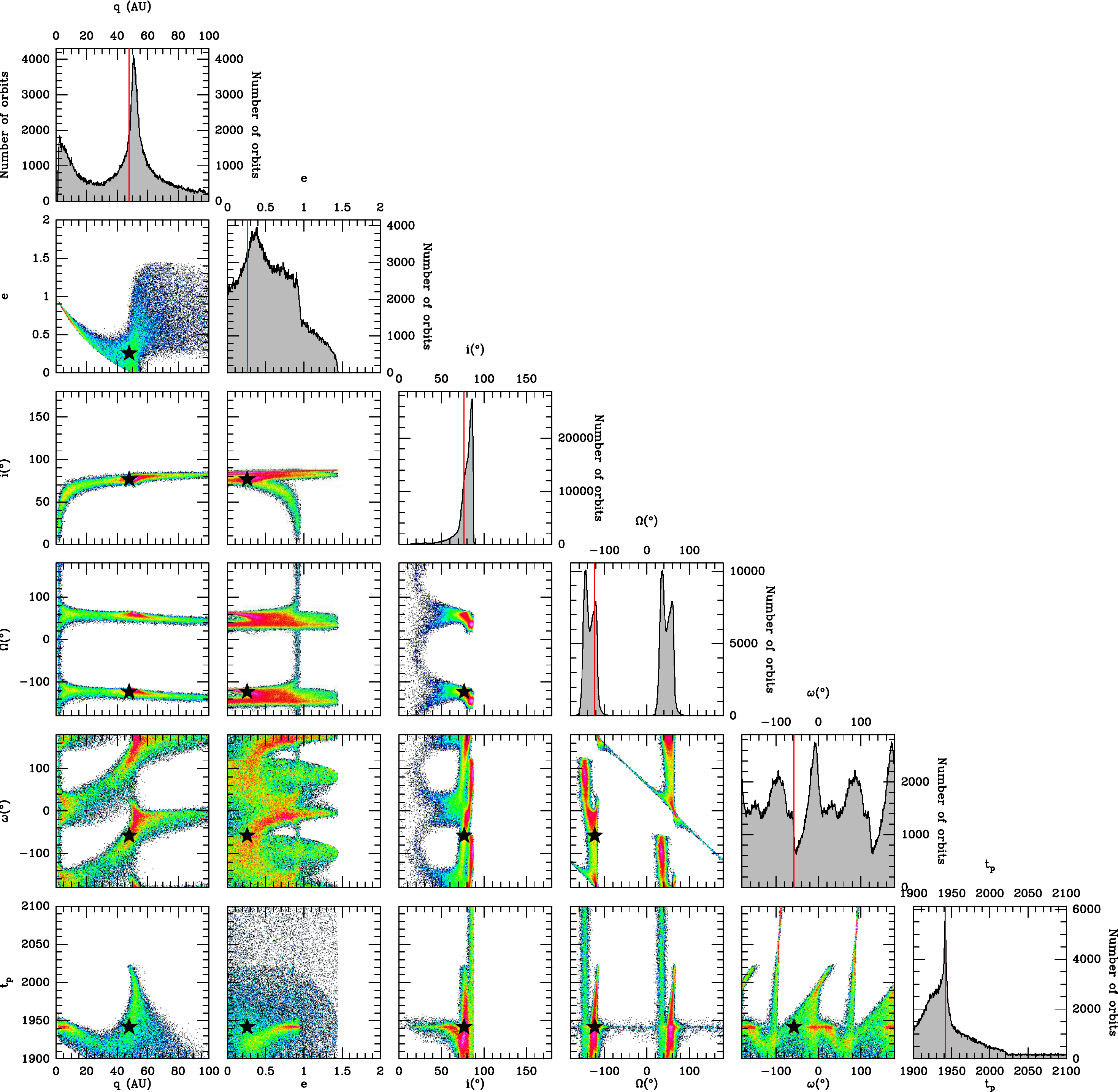}
      \caption{
      Posterior distributions of orbital elements of the close companion in \CHX\ using the universal Keplerian variable code by \citet{Beust2016}, with the possibility of unbound orbits ($e>1$). In all plots, the red bars (for histograms) and the black stars (in correlation maps) stand for the best-$\chi^2$ solution of the posterior distribution. For compatibility with unbound solutions, the first variable fitted is the periastron $q$ instead of the semi-major axis $a$. The $180\degr$ periodicity of distributions of the longitude of ascending node ($\Omega$) and the argument of periastron ($\omega$) is a degeneracy inherent to the astrometric fit with no radial velocity data.}
         \label{fig:corner_chx_unbound}
   \end{figure*}

We present in Fig.~\ref{fig:corner_chx_unbound} the orbital fitting results for \CHX\ using the universal Keplerian variable code by \citet{Beust2016}. This code allows for assessing unbound orbits in contrast to the analysis shown in \ref{sec:astrometry}. 

  \section{SPHERE and ALMA data overlay}
  
\begin{figure*}
   \centering
   \includegraphics[width=\hsize]{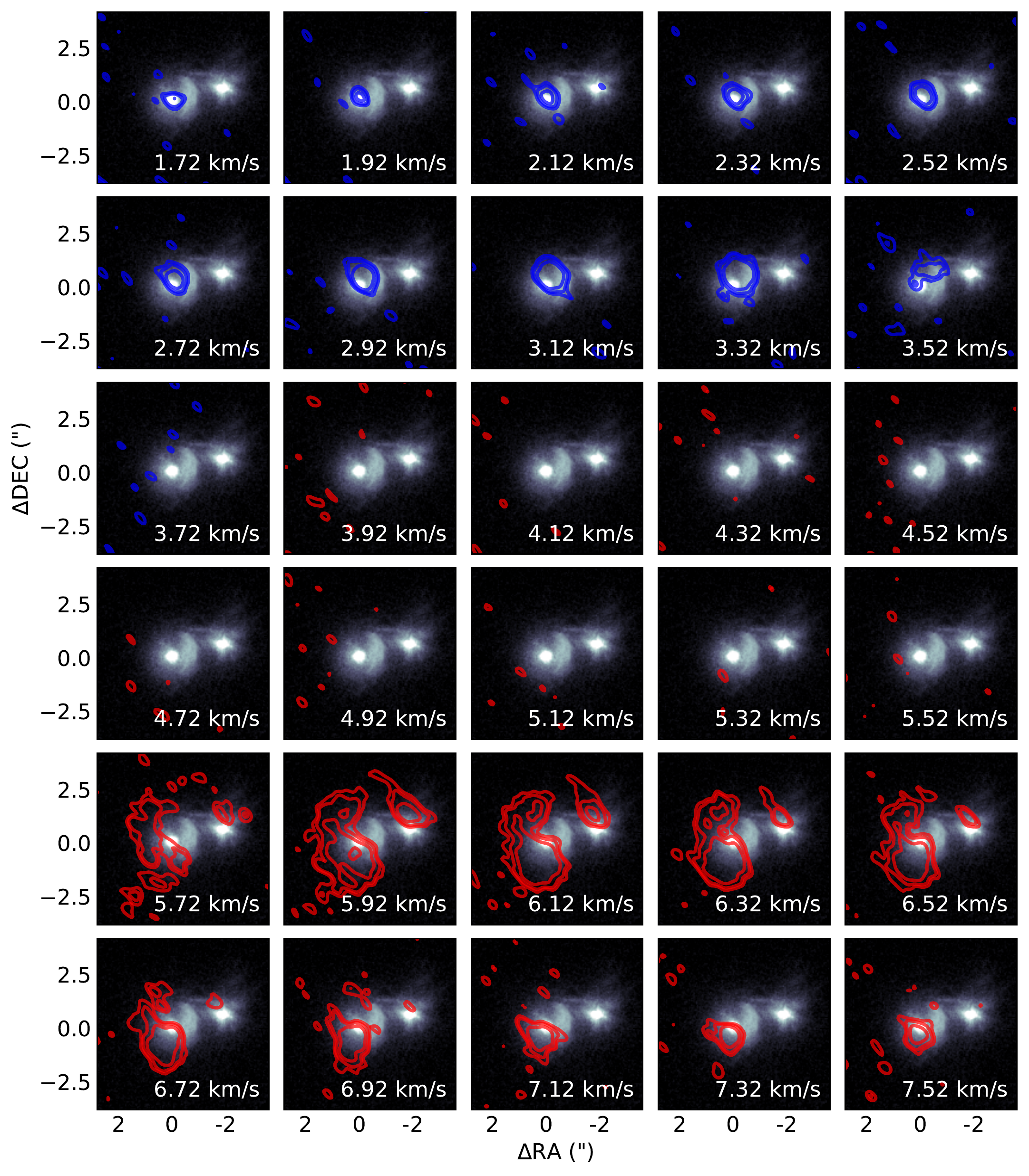}
      \caption{
      SPHERE PI image of \HPCHA\ with the ALMA CO $J=3-2$ channel maps as contours. The contours delineate the flux level of [3, 5, 10, 50]$\sigma$, where $\sigma$ is the standard deviation as measured in the image background. 
      The contours of channels redshifted relative to the systemic velocity ($\sim$3.8 \kms) are shown in red, and blueshifted channels are shown in blue.}
         \label{fig:hpcha_alma_channel}
   \end{figure*}

  We present in Fig.~\ref{fig:hpcha_alma_channel} the overlay of the scattered-light image and channel maps of ALMA CO emission for \HPCHA. The streamer feature seen in scattered light can also be recognized in the gas emission in redshifted channels.
  
\section{Coronagraphic images of \CHX, \SCRA, and \HPCHA} \label{app:stokes}
\begin{figure*}
   \centering
   \includegraphics[width=\hsize]{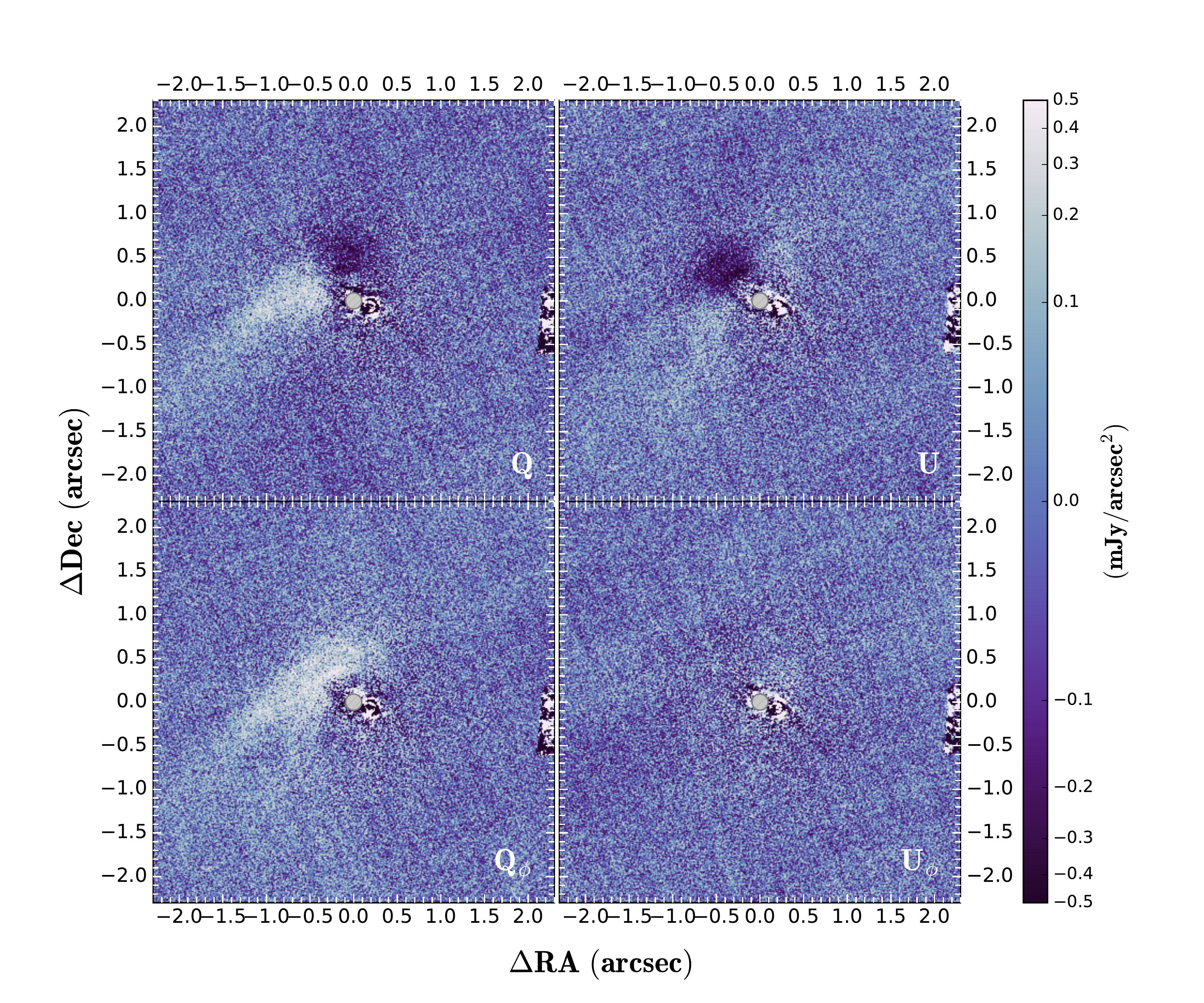}
      \caption{Stokes $Q$, $U$, $Q_\phi$, and $U_\phi$ polarized flux images of \CHX\ derived from SPHERE/IRDIS observations. The size of the coronagraphic mask is indicated with the grey, hashed circle.}
         \label{fig:chx_app}
   \end{figure*}

\begin{figure*}
   \centering
   \includegraphics[width=\hsize]{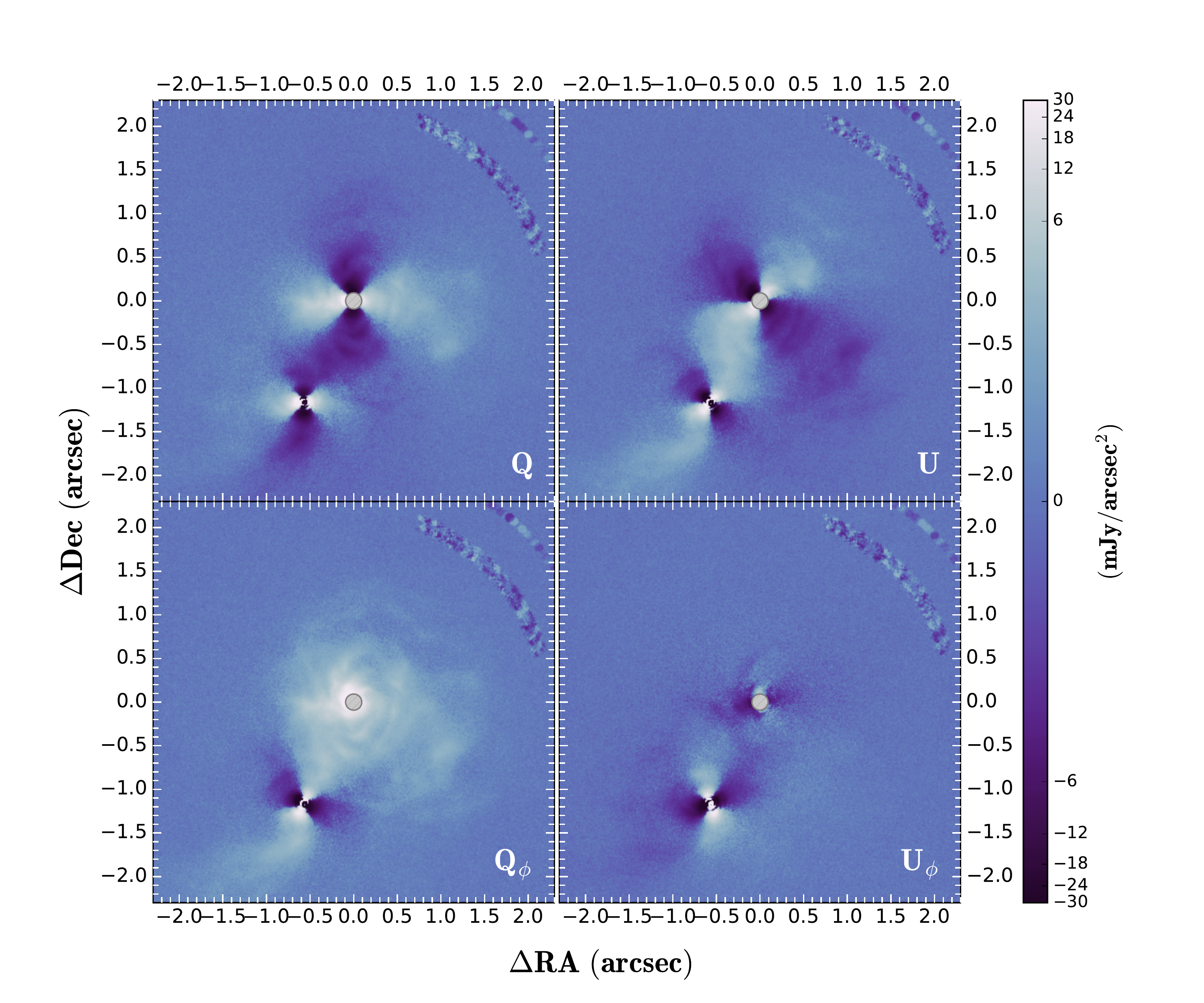}
      \caption{Similar as Fig.~\ref{fig:chx_app}, but for \SCRA.}
         \label{fig:scra_app}
   \end{figure*}

\begin{figure*}
   \centering
   \includegraphics[width=\hsize]{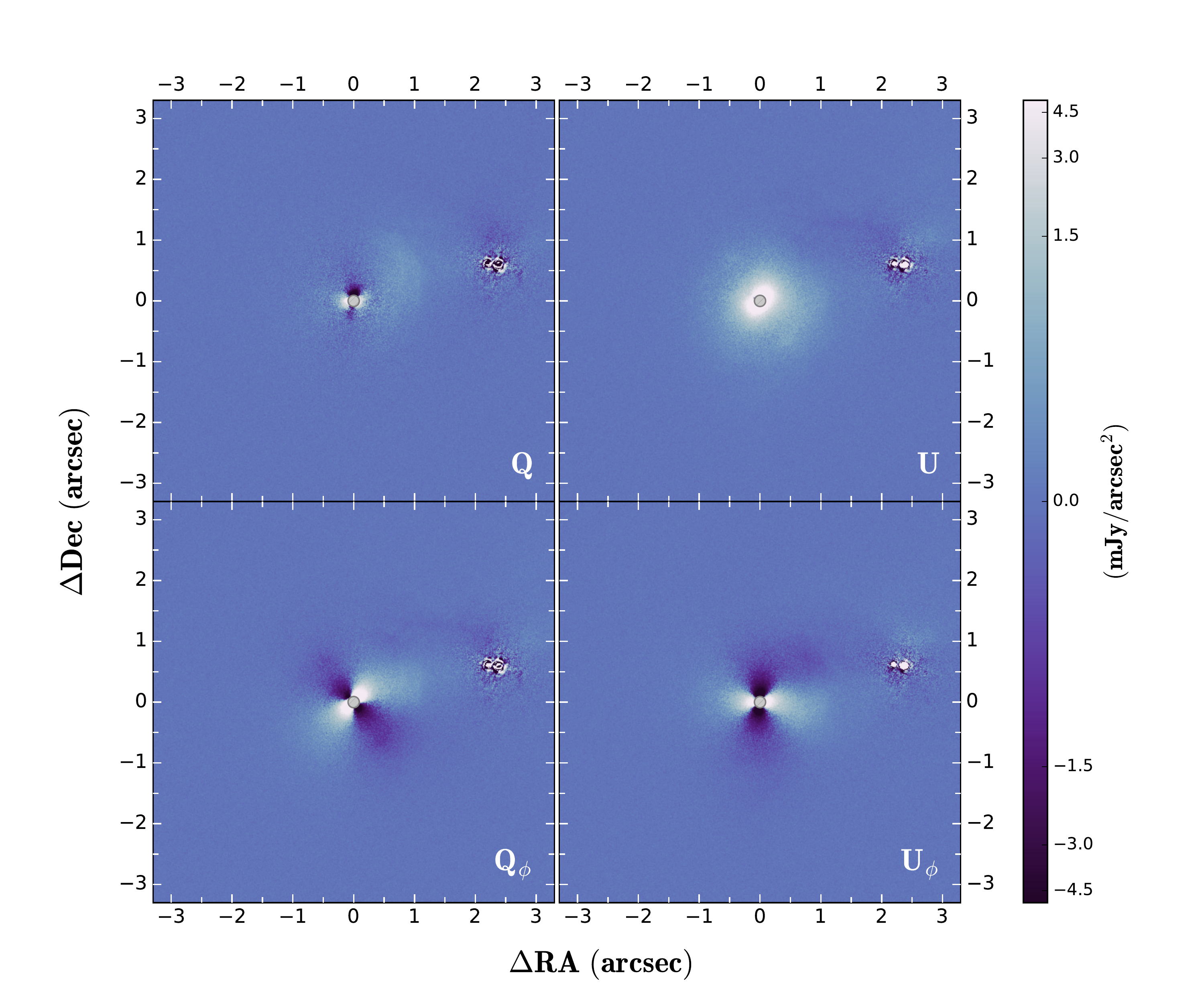}
      \caption{Similar as Fig.~\ref{fig:chx_app}, but for \HPCHA.}
         \label{fig:hpcha_app}
   \end{figure*}

We present the Stokes $Q$, $U$, $Q_\phi$, and $U_\phi$ polarized flux images obtained from SPHERE/IRDIS observations of the three multiple systems in Fig.~\ref{fig:chx_app}, \ref{fig:scra_app}, and \ref{fig:hpcha_app}.

\end{document}